\documentclass[aps,prfluids,reprint,groupedaddress,floatfix]{revtex4-2}
\usepackage{amsmath,amssymb}
\usepackage{graphicx}
\usepackage{dcolumn}
\usepackage{bm}
\usepackage{tabularx}
\newcolumntype{L}[1]{>{\raggedright\arraybackslash}p{#1}}
\newcolumntype{C}[1]{>{\centering\arraybackslash}p{#1}}
\newcolumntype{R}[1]{>{\raggedleft\arraybackslash}p{#1}}
\newcolumntype{Y}{>{\centering\arraybackslash}X}

\begin{document}

\title{Scale dependence and cross-scale transfer of kinetic
energy in compressible hydrodynamic turbulence at moderate Reynolds numbers}

\author{Petr Hellinger} \email{petr.hellinger@asu.cas.cz}
\affiliation{Astronomical Institute, CAS, Bocni II/1401, CZ-14100 Prague, Czech
Republic}
\affiliation{Institute of Atmospheric Physics, CAS, Bocni II/1401,
CZ-14100 Prague, Czech Republic}

\author{Andrea Verdini}
\affiliation{Dipartimento di Fisica e Astronomia, Universit\`a degli Studi di Firenze Largo E. Fermi 2, I-50125 Firenze, Italy}
\affiliation{INAF -- Osservatorio Astrofisico di Arcetri, Largo E. Fermi 5, I-50125 Firenze, Italy}
\author{Simone Landi}
\affiliation{Dipartimento di Fisica e Astronomia, Universit\`a degli Studi di Firenze Largo E. Fermi 2, I-50125 Firenze, Italy}
\affiliation{INAF -- Osservatorio Astrofisico di Arcetri, Largo E. Fermi 5, I-50125 Firenze, Italy}
\author{Emanuele Papini}
\affiliation{Dipartimento di Fisica e Astronomia, Universit\`a degli Studi di Firenze Largo E. Fermi 2, I-50125 Firenze, Italy}
\affiliation{INAF -- Osservatorio Astrofisico di Arcetri, Largo E. Fermi 5, I-50125 Firenze, Italy}
\author{Luca Franci}
\affiliation{School of Physics and Astronomy, Queen Mary University of London, London E1 4NS, United Kingdom}
\affiliation{INAF -- Osservatorio Astrofisico di Arcetri, Largo E. Fermi 5, I-50125 Firenze, Italy}
\author{Lorenzo Matteini}
\affiliation{Department of Physics, Imperial College London, London, SW7 2AZ, United Kingdom}
\affiliation{INAF -- Osservatorio Astrofisico di Arcetri, Largo E. Fermi 5, I-50125 Firenze, Italy}

\begin{abstract}

We investigate properties of the scale dependence and cross-scale transfer of kinetic 
energy in compressible three-dimensional hydrodynamic turbulence, by means of two direct 
numerical simulations of decaying turbulence with initial Mach numbers $M = 1/3$ and $M = 1$, 
and with moderate Reynolds numbers, $R_{\lambda} \sim 100$.
The turbulent dynamics is analyzed using compressible and incompressible versions of 
the dynamic spectral transfer (ST) and the K\'arm\'an-Howarth-Monin (KHM) equations. 
We find that the nonlinear coupling leads to a flux of the kinetic energy 
to small scales where it is dissipated; at the same time, the reversible 
pressure-dilatation mechanism causes oscillatory exchanges between the kinetic and internal 
energies with an average zero net energy transfer.  
While the incompressible KHM and ST equations are not generally valid in the simulations,
their compressible  counterparts are well satisfied and describe, 
in a quantitatively similar way, the decay of the kinetic energy on large scales, 
the cross-scale energy transfer/cascade, the pressure dilatation, and the dissipation. 
There exists a simple relationship between the KHM and ST results through the inverse proportionality between the wave vector $k$ and the spatial separation length $l$ 
as $k l \simeq \sqrt{3} $. 
{ For a given time
the dissipation and pressure-dilatation terms are strong
on large scales in the KHM approach whereas the ST terms 
become dominant on small scales; this is owing to the complementary cumulative
behavior of the two methods.
The effect of pressure dilatation is weak  
when averaged over a period of its oscillations 
and may lead to a transfer of the kinetic energy from large to small scales
without a net exchange between the kinetic and internal energies.
}
Our results suggest that for large-enough systems there exists an inertial range 
for the kinetic energy cascade. This transfer is partly owing to the classical,
nonlinear advection-driven cascade and partly owing to the pressure dilatation-induced energy 
transfer. 
We also use the ST and KHM approaches to investigate
properties of the internal energy. The dynamic ST and KHM equations for the internal energy
 are well satisfied in the simulations but behave very differently with respect 
to the viscous dissipation. We conclude that  ST and KHM approaches should better 
be used for the kinetic and internal energies separately. 

\end{abstract}

\maketitle

\section{Introduction}

Fundamental problems of turbulence concern how the energy (and other quantities) 
is distributed on spatio-temporal scales, how it is transferred across scales
and exchanged among its different forms.
The current understanding of turbulence is mostly based
on the hydrodynamic model in the incompressible
limit \cite{fris95}, where the divergence of the velocity field
is taken zero and a constant density
is usually assumed. In this case, 
the spatial-scale decomposition of the kinetic energy
(per mass) may be characterized by the
 spectral density of the velocity field \cite{pao65,pope00}
and its evolution can be
analyzed using spectral transfer (ST) approaches
\cite{alexal05,mini11}.
Alternatively, one can look at the cross-correlations
of the velocity field or structure functions
(related to the power spectrum), via
the K\'arm\'an-Howarth-Monin (KHM)
equation \citep{kaho38,kolm41b,moya75}.
 Another
possibility is to use space-filtering (coarse graining)
of the velocity field \citep{germ92,eyal09}. 
These approaches may be used to quantitatively characterize 
the different turbulence processes, the injection/decay,
the cross-scale energy transfer,
and the dissipation. 
Moreover, they can be used to determine 
whether an inertial range exists, where the only relevant
process is the cross-scale energy transfer, and
if this transfer is cascade-like \cite{rich22}, i.e. if
the cross-scale energy transfer is dominated by
interactions between nearby scales.

Extension of the incompressible results to
the case of general, compressible fluids with
variable densities is not trivial \cite{bazh99,eydr18,prgi19}. It is not evident how
to characterize, in an analogous manner, the scale-distribution of the kinetic
energy when the density is not constant 
\cite{alui11,alui13,laial18,scgr19}. There are multiple
different density weighting methods for
the spectral, structure function, and coarse graining
approaches.
Furthermore, the compressibility introduces the pressure-dilatation
effect that couples the kinetic and internal energies
in a reversible manner (in contrast to the irreversible viscous dissipation).
The pressure-dilatation channel brings into question the existence of an inertial range for
the kinetic energy. 

{
Numerical simulation results of Refs.~\onlinecite{aluial12,prgi19} 
indicate that the pressure-dilatation induced energy exchanges
tend to be more important on large scales.
Ref.~\onlinecite{aluial12} shows that the strength of pressure-dilatation effect
decreases on small scale so that there can exist a range of scales
where the pressure-dilatation is negligible and the kinetic
energy conservatively cascades.
On the other hand, Ref.~\onlinecite{wangal18}
shows that the pressure-dilatation appear on small scales 
(and may lead to cross-scale transfer of the
kinetic energy).}

Since the kinetic and internal energies are coupled via the dissipation, as well as
through the pressure dilatation, one may consider the total (kinetic+internal) energy,
that is strictly conserved. Refs.~\onlinecite{gaba11,baga14} formulate 
the KHM equation  in the compressible case for the total energy. They,
however, assume that the system follows a given closure (isothermal or polytropic)
{ and they use the closure to derive
the KHM equation}. In particular, they manipulate the pressure-dilatation
term to cast it in a form of a cascade rate; it is unclear if all or only
a part of pressure-dilatation effects are present in such a system.

{ Here 
 we address the pressure-dilatation effect, its role in
the compressible HD turbulence and its characteristic scales
using two methods.}
We reexamine the KHM equation for the kinetic energy
in compressible HD and analyze results of direct compressible HD
numerical simulations.
 We compare these results with those of a simple ST approach
in both the incompressible and compressible approximations. We also look at the
properties of the internal energy and its scale decomposition and compare these
results with those of the kinetic energy.
The paper is organized as follows: in Sec.~\ref{simulation} 
we present an overview of two direct 3D HD simulations.
In Sec.~\ref{spectr} we present the ST Fourier method and
use it to analyze the simulation results. 
In Sec.~\ref{cascade} we rederive the KHM equation for the kinetic energy
and apply it to the simulations results; results of 
 the two methods in both incompressible and compressible approximations are compared.
In Sec.~\ref{inten} we test the scale decomposition
of the internal energy using the ST and KHM approaches.
Finally, in Sec.~\ref{discussion} we discuss the obtained results.

\section{Numerical simulations}
\label{simulation}
We employ a 3D pseudo-spectral compressible hydrodynamic code derived from the compressible
MHD code \citep{verdal15} based on P3DFFT library \citep{peku12} and FFTW3
 \citep{frjo05}.  The  code solves 
the compressible Navier-Stokes equations for the fluid density $\rho$,
 velocity $\boldsymbol{u}$, and the pressure $p$:
\begin{align}
\frac{\partial \rho}{\partial t}+ \boldsymbol{\nabla} \cdot (\rho \boldsymbol{u}) &= 0,
\label{density}\\
\frac{\partial (\rho\boldsymbol{u})}{\partial t}+ \boldsymbol{\nabla}\cdot (\rho \boldsymbol{u}\boldsymbol{u})
&=-\boldsymbol{\nabla}p
 +\boldsymbol{\nabla}\cdot\boldsymbol{\tau},
\label{velocity}
\end{align}
complemented with an equation for the temperature $T=p/\rho$
\begin{align}
\frac{\partial T}{\partial t}+ (\boldsymbol{u} \cdot \boldsymbol{\nabla}) T =& \frac{\alpha}{\rho} \Delta T +
\frac{\gamma-1}{\rho} \left(-p \theta + \boldsymbol{\Sigma}:\boldsymbol{\tau}\right)
\label{temperature}
\end{align}
where 
$\theta=\boldsymbol{\nabla}\cdot \boldsymbol{u} $ is the dilatation,
$\boldsymbol{\Sigma}=\boldsymbol{\nabla} \boldsymbol{u}$ is the stress tensor,
and $\boldsymbol{\tau}$ is the viscous stress tensor
 ($\tau_{ij}=\mu\left(\partial u_{i}/\partial x_{j}+\partial u_{j}/\partial x_{i}-2/3 \theta \delta_{ij}\right)$; 
here the dynamic viscosity $\mu$ is assumed to be constant), 
and
 $\alpha$ is the thermal diffusivity (we set $\alpha=\mu$ and $\gamma=5/3$).
The colon operator denotes the double contraction of second-order tensors,
$\boldsymbol{\mathrm{A}}:\boldsymbol{\mathrm{B}}=\sum_{ij}A_{ij} B_{ij}$. 

For the compressible Navier-Stokes equations (\ref{velocity})
one gets the following equation for the average kinetic energy
in a closed system
\begin{align}
\frac{\partial \ }{\partial t} \left\langle \frac{1}{2} \rho |\boldsymbol{u}|^{2}\right\rangle
= \langle p \theta \rangle - \langle  \boldsymbol{\tau} : \boldsymbol{\Sigma}  \rangle,
\label{conskin}
\end{align}
where
$\langle \bullet \rangle$ denotes spatial averaging over the domain (the simulation box).
The two terms at the rhs of Eq.~(\ref{conskin}) couple the kinetic energy to the internal one  
\begin{align}
\frac{\partial \ }{\partial t} \left\langle  \rho e\right\rangle
= -\langle p \theta \rangle + \langle  \boldsymbol{\tau} : \boldsymbol{\Sigma}  \rangle
\label{consint}
\end{align}
where $e=T/(\gamma-1)$ is the internal energy density (per mass).

We perform two simulations of decaying turbulence with different levels of compressibility.
The simulation box size is $(2\pi)^3$ (with a grid of $1024^3$ points), 
periodic boundary conditions are assumed.
Both simulations are initialized with isotropic, random-phase, solenoidal 
fluctuations (i.e., $\theta$ is set to $0$)
on large scales (with wave-vector magnitudes $k=|\boldsymbol{k}|\leq 4$).
Run 1 starts with 
the rms Mach number $M=1/3$, whereas for run~2 we
set the initial Mach number $M=1$. 
For run~1 
 we set the (constant) dynamic viscosity 
$\mu=4\ 10^{-4}$, for run~2 we set
$\mu=2\ 10^{-3}$; we use a large viscosity in
this case to avoid steep gradients (shocks)
that are not well resolved by pseudo-spectral
codes. 
Table~\ref{tab} gives an overview of the simulation parameters.
Table~\ref{tab} also shows the times $t_\omega$ where
the rms of the vorticity reaches the maximum and
the microscale Reynolds number, $R_{\lambda}$, given by \cite{kior92}
\begin{align}
R_{\lambda} = \left(\frac{5}{3 \langle \boldsymbol{\tau} : \boldsymbol{\Sigma} \rangle  }  \right)^{1/2} 
\left\langle \rho  \right\rangle \left \langle |\boldsymbol{u}|^2\right\rangle 
\end{align}
at that time.

\begin{table}
\caption{Simulation parameters}
\label{tab}
\begin{tabularx}{0.4\textwidth}{c|YYYYYY} 
run & grid & size & $M$ & $\mu$ & $t_{\omega}$ & $R_{\lambda}$  \tabularnewline
\hline
1 & $1024^{3}$ & $(2\pi)^{3}$ & $1/3$ &  $4\ 10^{-4}$ & 6.6  & 146 \tabularnewline
2 & $1024^{3}$ & $(2\pi)^{3}$ & 1 & $2\ 10^{-3}$ & 6.5  & 82  \tabularnewline
\end{tabularx}
\end{table}

\begin{figure}
\centerline{\includegraphics[width=8cm]{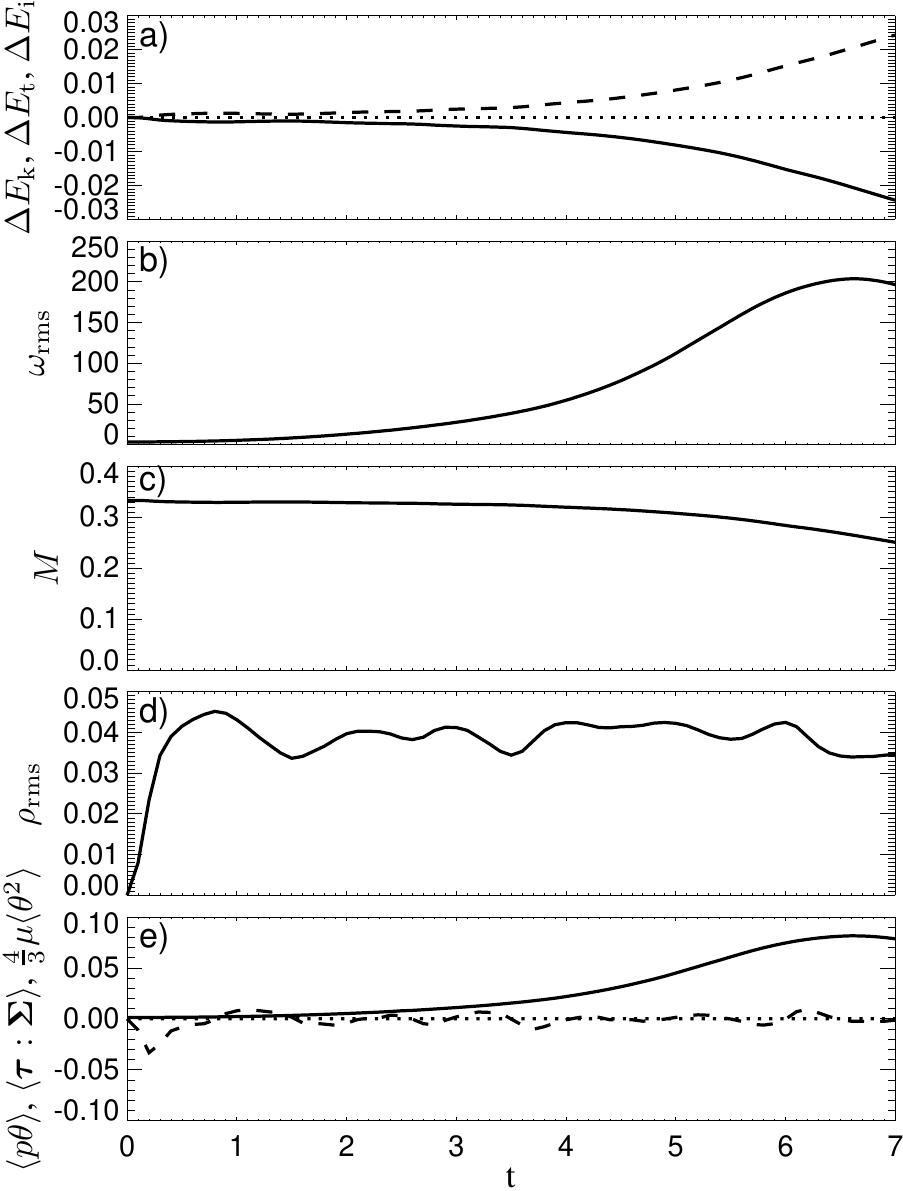}}
\caption{Evolution in run~1: (a) the relative changes
in the kinetic energy $\Delta E_\mathrm{k}$ (solid line), the total energy
 $\Delta E_\mathrm{t}$ (dotted line), and the internal energy $\Delta E_\mathrm{i}$ (dashed),
(b) rms of the vorticity $\omega_\text{rms}$, (c) Mach number $M$,
(d) rms of the density fluctuations $\rho_\text{rms}$, and
(e) the dissipation rate $\langle \boldsymbol{\tau}:\boldsymbol{\Sigma}\rangle$ (solid line),
the pressure-dilatation rate $\langle p\theta \rangle$ (dashed line), and
the compressible dissipation rate $4 \mu \langle \theta^2 \rangle/3$ (dotted line)
 as functions of time.
\label{evolr1}
}
\end{figure}

The evolution of run~1 is shown in Fig.~\ref{evolr1}. In this simulation the
total energy $E_\mathrm{t}=E_\mathrm{k}+E_\mathrm{i}$ is well conserved. 
Here $E_\mathrm{k}=\langle \rho u^2 \rangle/2$ is the kinetic energy and
$E_\mathrm{i}=\langle \rho e \rangle$ is the internal one.
Fig.~\ref{evolr1}a displays the evolution of the relative changes in these energies,
$\Delta E_\mathrm{k,i,t}=(E_\mathrm{k,i,t}(t)-E_\mathrm{k,i,t}(0))/E_\mathrm{t}(0)$
(the solid line denotes the kinetic energy, the dashed line the internal one,
and the dotted line denotes the total energy).
 The relative change
of the total energy is negligible, $|\Delta E_\mathrm{t}(t=7)| < 8\, 10^{-6}$,
the kinetic energy is transformed to the internal one.
Fig.~\ref{evolr1}b shows the evolution
of the rms of the vorticity
$\boldsymbol{\omega}=\boldsymbol{\nabla}\times \boldsymbol{u}$,
$\omega_\text{rms}^2= \langle |\boldsymbol{\omega}|^2\rangle$.
{
The vorticity reaches the maximum of about $200$ at $t=t_{\omega}\simeq 6.6$; 
this corresponds to the maximum of the (incompressible) dissipation rate and 
may be considered as a signature of a fully developed turbulent cascade
in a decaying system. After about this time the dissipation rate of 
the kinetic energy varies only slowly indicating a quasi-stationary evolution 
\cite{pouqal10}.}
Fig.~\ref{evolr1}c
displays the evolution of the average Mach number $M$ (i.e., the
ratio between rms of the velocity and the mean sound speed). $M$ slowly decreases
during the evolution due to the decay of the level of fluctuations as
well as due to the turbulent heating that leads to
an increasing sound speed.
Fig.~\ref{evolr1}d shows the rms of the density fluctuations, 
$\rho_\text{rms}^2=\langle (\rho -\rho_0)^2\rangle$
(where $\rho_0=\langle \rho \rangle$). Weak fluctuations ($\rho_\text{rms}\sim 0.04$) 
develops during the first phase of the relaxation of the initial,
constant $\rho$ conditions. 
Fig.~\ref{evolr1}e quantifies the evolution
of the dissipation rate $\langle  \boldsymbol{\tau} : \boldsymbol{\Sigma}  \rangle$
 (solid line), the pressure dilatation term $\langle p \theta \rangle$
(dashed line), and the compressible dissipation term $4 \mu \langle \theta^2 \rangle/3$
(dotted line). In run~1, the compressible dissipation is negligible,
the dissipation rate follows closely the behavior of the vorticity (see Fig.~\ref{evolr1}b).
A relatively large pressure dilatation rate develops initially as
a relaxation of the initial solenoidal conditions. At later times 
the pressure dilatation becomes weaker than the dissipation rate
and oscillates around zero \cite{papo87,kior92}. Taking an average over about
a period of these oscillations removes 
the exchange between the kinetic and internal energies induced by the pressure dilatation,
$\langle \langle p \theta \rangle \rangle_t\simeq 0$;
henceforth $\langle \bullet \rangle_t$ denotes time averaging.

\begin{figure}
\centerline{\includegraphics[width=8cm]{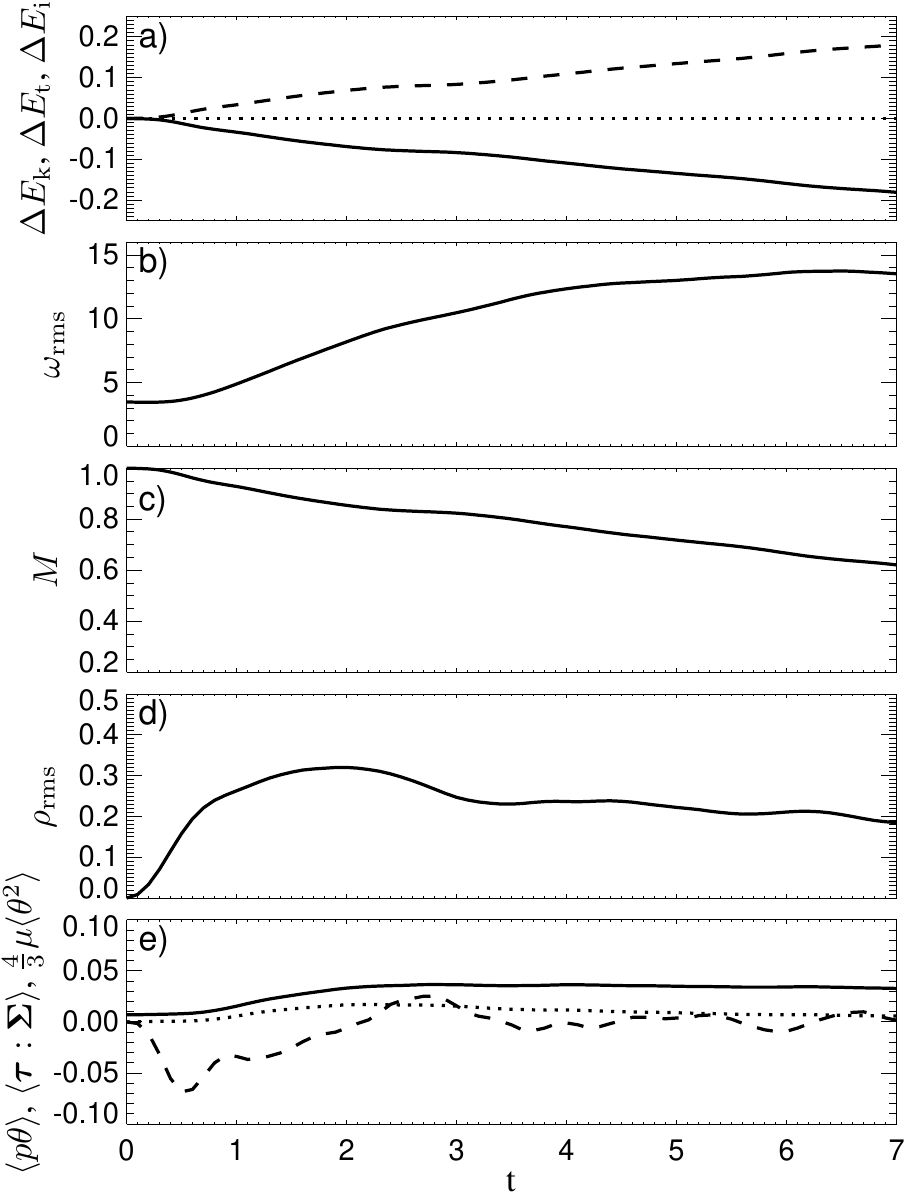}}
\caption{Evolution in run~2: (a) the relative changes
in the kinetic energy $\Delta E_\mathrm{k}$ (solid line), the total energy
 $\Delta E_\mathrm{t}$ (dotted line), and the internal energy $\Delta E_\mathrm{i}$ (dashed),
(b) rms of the vorticity $\omega_\text{rms}$, (c) Mach number $M$,
(d) rms of the density fluctuations $\rho_\text{rms}$, and
(e) the dissipation rate $\langle \boldsymbol{\tau}:\boldsymbol{\Sigma}\rangle$ (solid line)
the pressure-dilatation rate $\langle p\theta \rangle$ (dashed line), and
the compressible dissipation rate $4 \mu \langle \theta^2 \rangle/3$ (dotted line)
 as functions of time.
\label{evolr2}
}
\end{figure}

Run~2 exhibits an evolution  qualitatively similar to that of run~1
as shown in Fig.~\ref{evolr2}. Fig.~\ref{evolr2}a
displays the evolution of the relative changes in the kinetic,
internal, and  total energies; the relative change
of the total energy in run~2 is also negligible, $|\Delta E_\mathrm{t}(t=8)| < 4\, 10^{-5}$.
Fig.~\ref{evolr2}b shows that
the rms of vorticity in run~2 reaches the maximum of about $14$ at $t=t_{\omega}\simeq 6.5$.
This is much smaller that in run~1 likely due to the larger viscosity and compressibility.
In run~2, the Mach number (Fig.~\ref{evolr2}c) decreases faster then in run~1
since the turbulent heating leads to larger relative changes of the
temperature for the colder fluid.
The larger Mach number leads to important density variations,
Fig.~\ref{evolr2}d shows that the rms of the density rapidly
becomes about $0.3$. For later times $\rho_\text{rms}$ tends
to slowly decrease.   
 Fig.~\ref{evolr2}e shows the properties of dissipation and the pressure dilatation.
In run~2, the compressible dissipation is not negligible and, especially
during the initial phase, compressible dissipation makes an important fraction
of the total dissipation. The dissipation rate is interestingly smaller
in run~2 compared to run~1 whereas the pressure dilatation is more
important.
{ 
The kinetic energy decreases overall with time whereas the internal
energy increases owing to the viscous dissipation. On top of
this trend both the energies
exhibit noticeable oscillations owing to the pressure dilatation-induced exchanges;
these oscillations are also seen in the vorticity, and the Mach number.}
As in run~1 the pressure dilatation at later times oscillate around
zero and disappear when time-averaged.

\begin{figure}
\centerline{\includegraphics[width=8cm]{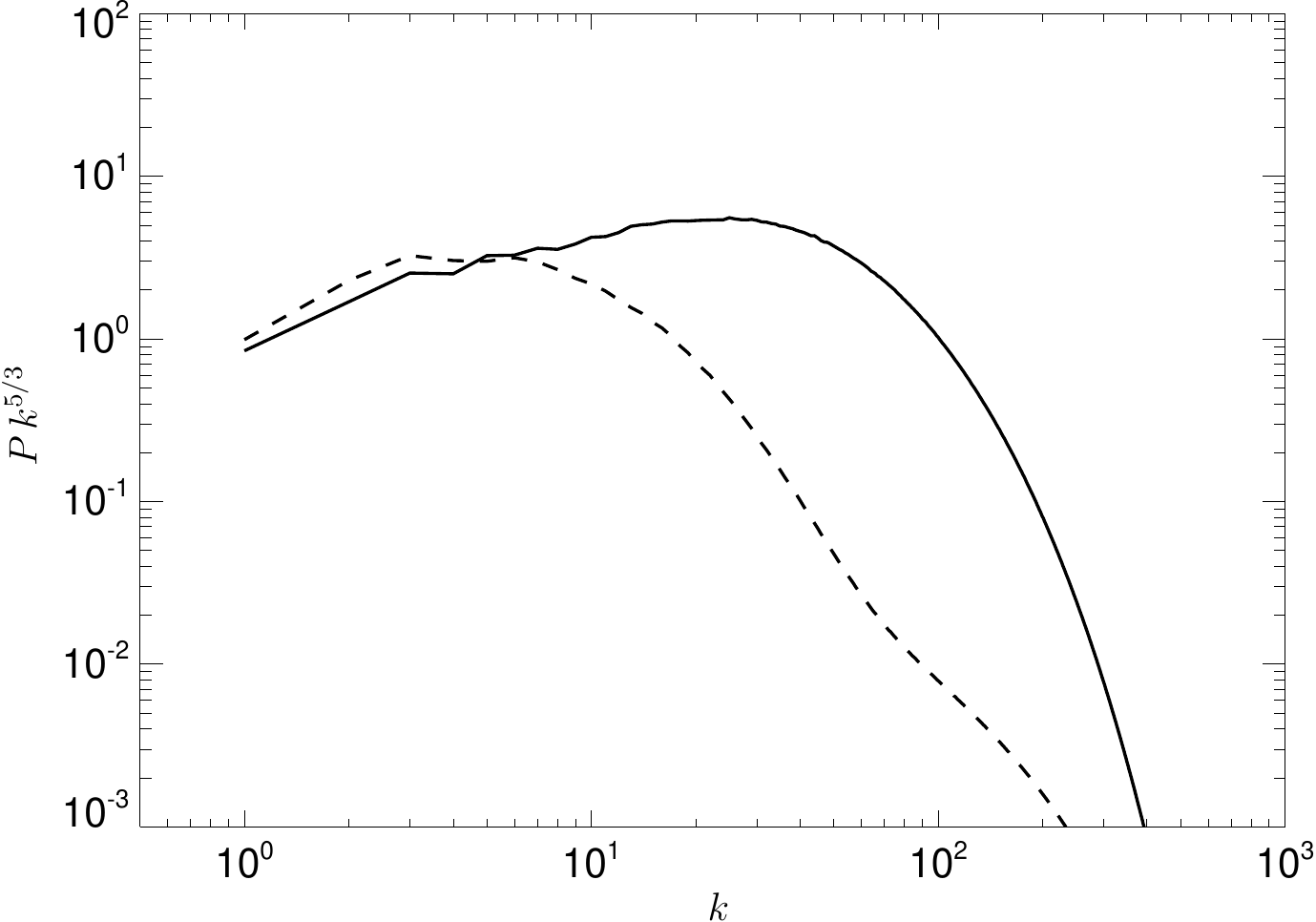}}
\caption{Power spectral density of $\boldsymbol{u}$,
compensated by $k^{5/3}$
as a function of the wave vector $k$
in run~1 (solid line) and run~2 (dashed line).
\label{spec}
}
\end{figure}

Fig.~\ref{spec} shows the power spectral density (PSD) compensated by $k^{5/3}$
of the velocity fluctuation at the time $t_{\omega}$,
  $t_{\omega} \simeq 6.6$ for run~1 and  $t_{\omega}\simeq 6.5$ for run~2.
Both the PSDs exhibit hints of the Kolmogorov-like scaling,
only a very small range of wavevectors have slopes compatible with $-5/3$ 
($k\in[12,30]$ and $k\in[3,6]$ for run~1 and~2, respectively)
 prior to the steepening due to the dissipation. 
The analysis of the energy transfer will show that only these scales 
can be roughly identified as the inertial range.

\section{Spectral Transfer}

 \label{spectr}

\subsection{Incompressible HD}
We start with the incompressible Navier-Stokes equation
\begin{align}
\frac{\partial\boldsymbol{u}}{\partial t}+ (\boldsymbol{u}\cdot \boldsymbol{\nabla})\boldsymbol{u}
&=-\frac{\boldsymbol{\nabla}p}{\rho}
 + \nu \Delta \boldsymbol{u},
\label{ivelocity}
\end{align}
where $\boldsymbol{u}$ is the velocity field, $\rho$ the density, $p$ the pressure,
$\nu$ is the kinematic viscosity. 
Beside the incompressibility, $\theta=0$, we also
assume that the density is constant $\rho=\rho_0$; henceforth we set $\rho_0=1$. 
In this system the equation for the kinetic energy (per mass) reads
\begin{align}
\frac{\partial \ }{\partial t} \left\langle \frac{1}{2} |\boldsymbol{u}|^{2}\right\rangle
=-\nu \left\langle\boldsymbol{\nabla}\boldsymbol{u}:\boldsymbol{\nabla}\boldsymbol{u} \right\rangle
=- \epsilon
\label{icons}
\end{align}
where $\epsilon$ is  the incompressible
dissipation rate (per mass).

Taking the Fourier transform of Eq.~(\ref{ivelocity}) one gets an equation 
for the amplitude of a given Fourier mode
$$\widehat{\boldsymbol{u}} (\boldsymbol{k})= \sum_{\boldsymbol{x}} \boldsymbol{u}(\boldsymbol{x})
 \mathrm{exp}(i \boldsymbol{k} \cdot \boldsymbol{x} )  
$$
\begin{align}
\frac{1}{2}\frac{\partial|\widehat{\boldsymbol{u}}|^2}{\partial t}+ 
\Re \widehat{\boldsymbol{u}}^* \cdot \widehat{(\boldsymbol{u}\cdot \boldsymbol{\nabla})\boldsymbol{u}}
&= - \nu k^2 |\widehat{\boldsymbol{u}}|^2,
\label{st1m}
\end{align}
where wide hats denote the Fourier transform,
asterisks signify the complex conjugate, and $\Re$ means the real part.

For the kinetic energy in modes with wave-vector magnitudes smaller than 
or equal to $k$
(we take a low-pass filter in the Fourier space)
\begin{align}
E_{\mathrm{k}k}^{(i)} &= \frac{1}{2} \sum_{|\boldsymbol{k}^\prime|\le k  } 
|\widehat{\boldsymbol{u}}(\boldsymbol{k}^\prime)|^2 
\end{align}
one gets this dynamic equation 
\begin{align}
\frac{\partial E_{\mathrm{k}k}^{(i)} }{\partial t} +  S_k^{(i)} = - D_k^{(i)}
\label{isptrdyn}
\end{align}
where
\begin{align}
S_k^{(i)} &= \Re \sum_{|\boldsymbol{k}^\prime|\le k}
\widehat{\boldsymbol{u}}^*(\boldsymbol{k}^\prime)\cdot 
\widehat{[(\boldsymbol{u}\cdot \boldsymbol{\nabla})\boldsymbol{u}]}(\boldsymbol{k}^\prime)    
 \\
D_k^{(i)} &= \nu \sum_{|\boldsymbol{k}^\prime|\le k} |\boldsymbol{k}^\prime|^2 |\widehat{\boldsymbol{u}}(\boldsymbol{k}^\prime)|^2 
\end{align}
Henceforth the superscript $(i)$ denotes the incompressible approximation.
In Eq.~(\ref{isptrdyn}) $S_k^{(i)}$ describes the energy transfer (cascade) to scales with
wave-vector magnitudes larger than $k$
whereas $D_k^{(i)}$ signifies the viscous
dissipation on scales with wave-vector magnitudes smaller than or equal to $ k$.
Eq.~(\ref{isptrdyn}) may also serve to determine 
the inertial range as a region where 
\begin{align}
  S_k^{(i)} = \epsilon,
\label{iexactST}
\end{align}
i.e., where the energy transfer/cascade rate equals to the dissipation one.

\subsection{Compressible HD}

To characterize the spectral decomposition of the kinetic
energy in the compressible case we define 
the density-weighted velocity field \cite{kior90}
\begin{align}
\boldsymbol{w}=\rho^{1/2} \boldsymbol{u}.
\end{align}
Taking the Fourier transform of Eq.~(\ref{velocity}) one gets an equation
for an amplitude of a given Fourier mode $\widehat{\boldsymbol{w}}(\boldsymbol{k})$
as \cite{scgr19,prgi19}
\begin{align}
\frac{1}{2}\frac{\partial|\widehat{\boldsymbol{w}}|^2}{\partial t}=&
-\Re \widehat{\boldsymbol{w}}^* \cdot \widehat{(\boldsymbol{u}\cdot \boldsymbol{\nabla})\boldsymbol{w}}
- \frac{1}{2} \Re \widehat{\boldsymbol{w}}^* \cdot \widehat{\theta \boldsymbol{w}} 
\label{cst1m}\\
& - \widehat{\boldsymbol{w}}^* \cdot \widehat{ \rho^{-1/2} \boldsymbol{\nabla} p} 
   + \widehat{\boldsymbol{w}}^* \cdot \widehat{ \rho^{-1/2} \boldsymbol{\nabla}\cdot \boldsymbol{\tau}}
\nonumber
\end{align}

For the kinetic energy in modes with wave-vector magnitudes smaller than
or equal to $k$ 
\begin{align}
E_{\mathrm{k}k} &= \frac{1}{2} \sum_{|\boldsymbol{k}^\prime|\le k  }
|\widehat{\boldsymbol{w}}(\boldsymbol{k}^\prime)|^2 
\end{align}
one gets, analogously to the incompressible case, the following equation
\begin{align}
\frac{\partial E_{\mathrm{k}k}}{\partial t} +  S_k = \Psi_k - D_k
\label{sptrdyn}
\end{align}
where (henceforth we will drop the $\boldsymbol{k}^\prime$ argument)
\begin{align}
S_k &= \Re \sum_{|\boldsymbol{k}^\prime|\le k}
\widehat{\boldsymbol{w}}^*
\cdot
\widehat{(\boldsymbol{u}\cdot \boldsymbol{\nabla})\boldsymbol{w}}
+ \frac{1}{2} \Re \sum_{|\boldsymbol{k}^\prime|\le k}
\widehat{\boldsymbol{w}}^*
\cdot
\widehat{\theta \boldsymbol{w}}
\label{stsk}
\\
\Psi_k &=- \Re \sum_{|\boldsymbol{k}^\prime|\le k}
\widehat{\boldsymbol{w}}^*
\cdot \widehat{ \rho^{-1/2} \boldsymbol{\nabla} p  }
 \\
D_k &=-\Re \sum_{|\boldsymbol{k}^\prime|\le k} 
\widehat{\boldsymbol{w}}^* 
\cdot
\widehat{\rho^{-1/2} \boldsymbol{\nabla} \cdot \boldsymbol{\tau}}.
\end{align}
Here $S_k$ represents the energy transfer/cascade rate, $\Psi_k$ describes the pressure-dilatation effect,
and $D_k$ is the dissipation rate for modes with wave-vector magnitude smaller than or equal to
 $k$.
For large wave vectors, one gets unfiltered values 
\begin{align}
E_{\mathrm{k}k}\rightarrow E_{\mathrm{k}}, \ \ \
\Psi_k \rightarrow \langle p\theta \rangle, \ \ \text{and}   \  \
D_k  \rightarrow Q_\mu, 
\label{stcum}
\end{align}
where $Q_\mu$ is the viscous dissipation rate, $Q_\mu=\langle  \boldsymbol{\tau} : \boldsymbol{\Sigma}  \rangle$.
The inertial range could be defined as 
\begin{align}
S_k= Q_\mu
\label{exactST}
\end{align}
but this equation neglects the pressure dilatation.

\begin{figure}
\centerline{\includegraphics[width=8cm]{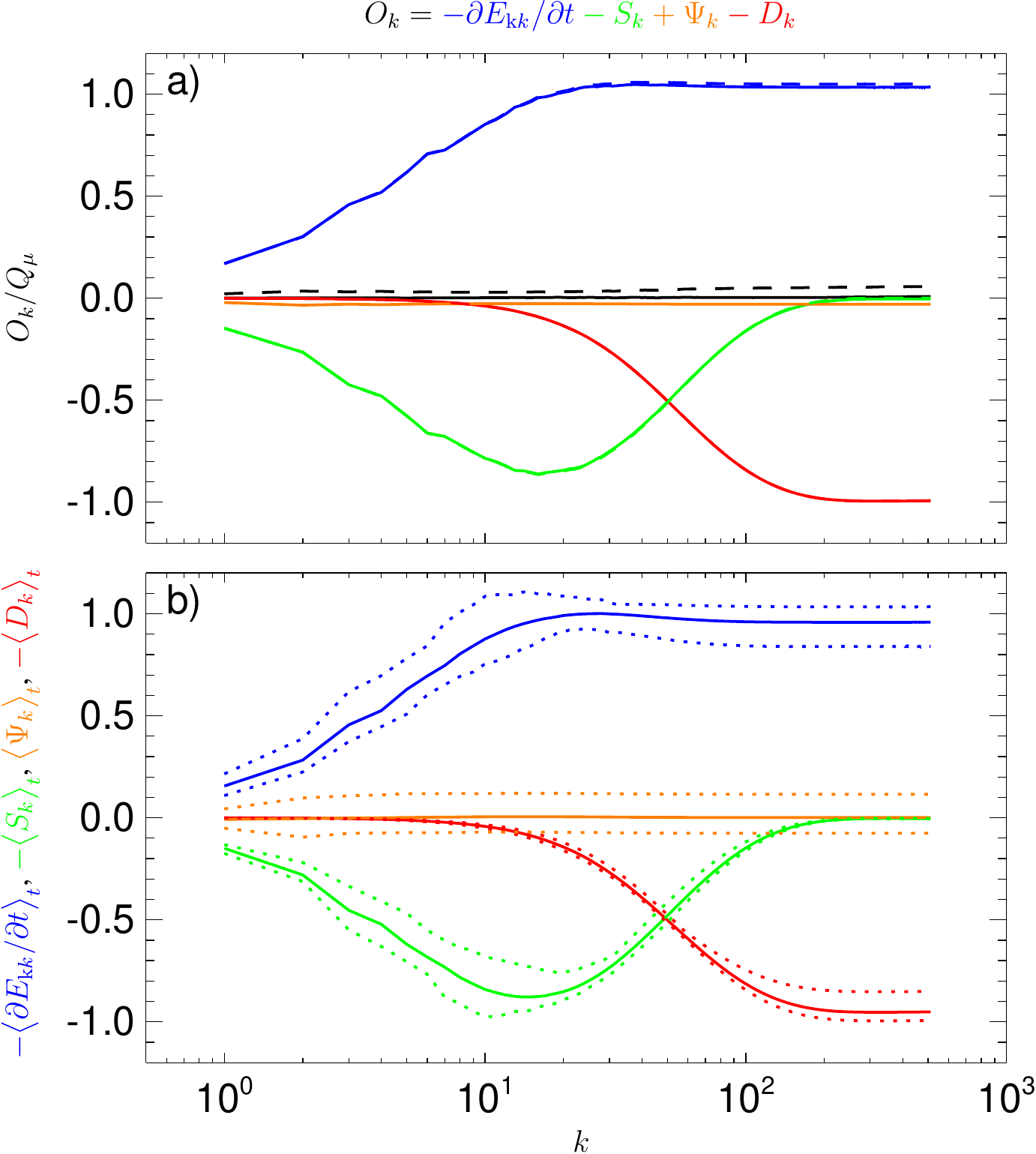}}
\caption{Spectral transfer in run~1. (a)  The validity test  $O_{k}$
 of Eq.~(\ref{ok}) (black solid line)
 as a function of $k$ along with
the different contributions (solid lines): (blue) the losses/decay
term $-\partial E_{\mathrm{k}k}/{\partial t}$, (green)
the transfer term $-S_k$, (orange) the pressure dilatation term
 $ \Psi_k$, and (red) the dissipation term $-D_{k}$.
Dashed lines show the incompressible equivalent $O_k^{(i)}$ (Eq.~(\ref{iok}))
and its contributions.
(b) Time-averaged contributing terms (solid lines) 
with their minimum and maximum values (dotted lines) for 
(blue) the decay $-\langle\partial E_{\mathrm{k}k}/{\partial t}\rangle_t$,
(green) the transfer $-\langle S_k\rangle_t$, (orange) the pressure dilatation term
 $ \langle\Psi_k\rangle_t$, and (red) the dissipation term $ -\langle D_{k}\rangle_t $.
All the quantities are normalized with respect to $Q_{\mu}$.
\label{spectr1}
}
\end{figure}

To validate the conservation of energy at any given scale, 
expressed by Eqs.~(\ref{sptrdyn}) and Eq.~(\ref{isptrdyn}), 
and to compare the incompressible and compressible decomposition, 
we introduce the error, i.e., the departure from the conservation of energy, 
for the compressible case
\begin{align}
O_k = -\frac{\partial E_{\mathrm{k}k}}{\partial t} -  S_k + \Psi_k - D_k
\label{ok}
\end{align}
and for the incompressible case:
\begin{align}
O_k^{(i)} = -\frac{\partial E_{\mathrm{k}k}^{(i)} }{\partial t} -  S_k^{(i)} - D_k^{(i)}.
\label{iok}
\end{align}
Fig.~\ref{spectr1}a displays results 
of the spectral transfer analysis for run~1, solid lines show
(black) $O_k$ and its contributions
(blue) the rate of change/decaying
term $-\partial E_{\mathrm{k}k}/{\partial t}$,
(green) the energy transfer/cascade term $-S_k$, 
(orange) the pressure dilatation term 
 $ \Psi_k$, and (red) the dissipation term $-D_{k} $
 (all normalized with respect to $Q_\mu$) as functions of $k$. 
Dashed lines show the corresponding error of
the incompressible approximation $O_k^{(i)}$. 
and its contributions.
The validity tests $O_k$ and $O_k^{(i)}$ in Fig.~\ref{spectr1}a
are calculated at $t_\omega$ and $t_\omega+\Delta t$ with $\Delta t=0.1$, $\partial E_{\mathrm{k}k}/\partial t$
is approximated by the finite difference
$\partial E_{\mathrm{k}k}/\partial t \approx [E_{\mathrm{k}k}(t_\omega+\Delta t)-E_{\mathrm{k}k}(t_\omega)]/\Delta t$.
Eq.~(\ref{sptrdyn}) is well satisfied $|O_k|/Q_\mu <0.01$ the error is partly numerical,
likely related to the finite-difference approximation 
of $\partial E_{\mathrm{k}k}/\partial t$.
The rate of change of the kinetic energy
$\partial E_{\mathrm{k}k}/\partial t$ is negative and varies mostly
on large scales; the energy-containing range is then on large scales, roughly for
wave-vector magnitudes smaller than about 3.  
The spectral energy transfer/cascade rate
$S_k$ dominates on medium scales, with maximum around $k=20$. 
The viscous dissipation $D_k$ is important on small scales. 
The pressure dilatation is weak in this weakly compressible case;
the incompressible predictions are close to their compressible counterparts.
The error of the incompressible approach $O_k^{(i)}$ appears to be related
to the neglected pressure dilatation term.

In run~1, the pressure-dilatation effect is small but non-negligible at
a given time. As the pressure dilatation oscillates in time, 
it is interesting to look at
the time-averaged quantities in Eq.~(\ref{sptrdyn}).
Fig.~\ref{spectr1}b displays the different terms averaged over the time $5.8\div 7.0$ 
(during this period the system is quasi-stationary, see Fig.~\ref{evolr1})
by solid lines. The dotted lines show the corresponding maximum and minimum values.
There we see that even in the weakly compressible run~1,
the different terms fluctuate with an important amplitude (of the order of $0.1  Q_\mu$).
However, the pressure dilatation is, on average, negligible on all scales. 
Finally we note that
in run~1 there is no inertial range as $S_k$ reaches maximally about $0.8 Q_\mu$
in a region where the dissipation is not negligible.

\begin{figure}
\centerline{\includegraphics[width=8cm]{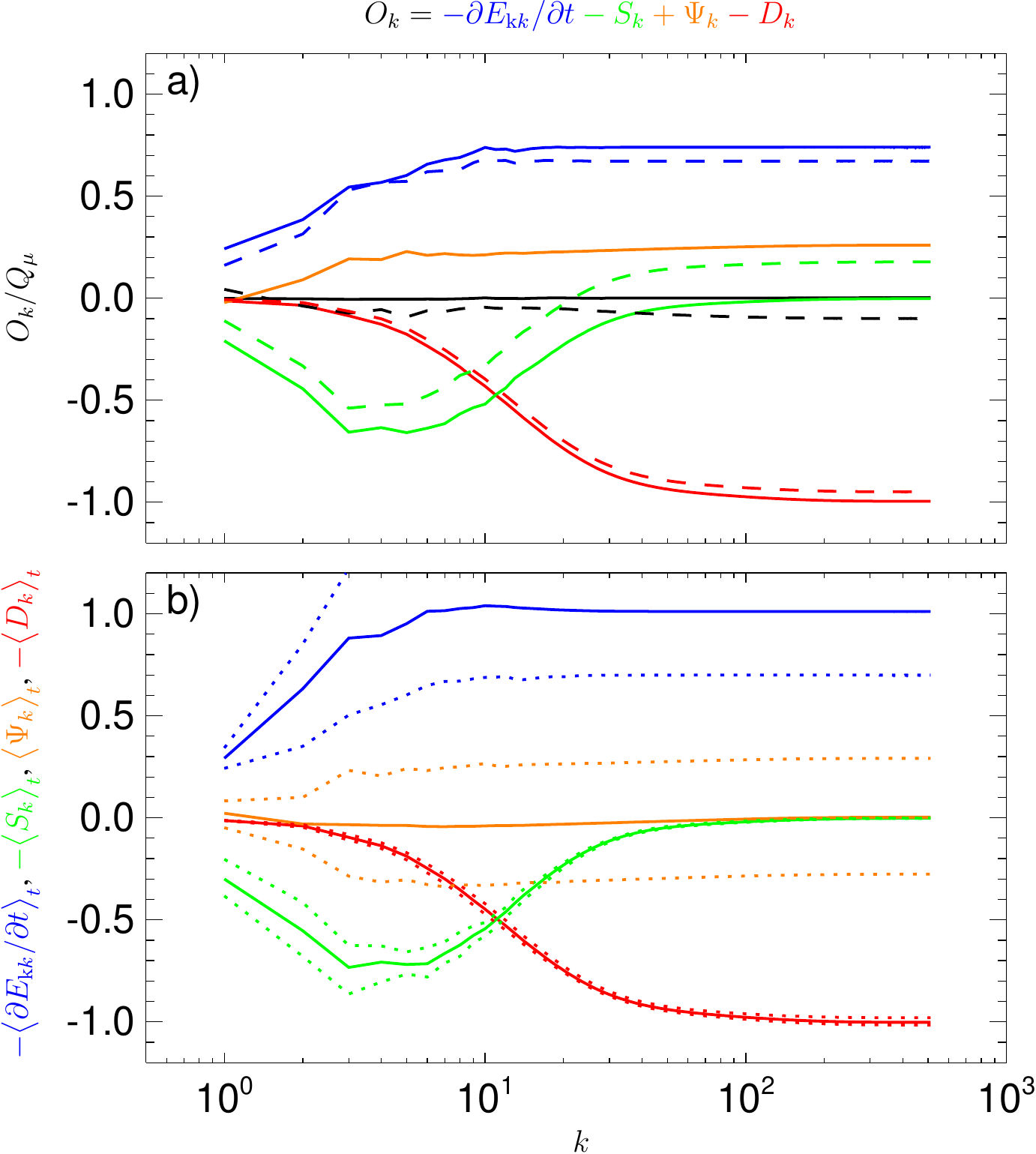}}
\caption{Spectral transfer in run~2 in the same
format as in Fig.~\ref{spectr1}.
\label{spectr2}
}
\end{figure}

Results from the ST approach in run~2 are shown in 
Fig.~\ref{spectr2} in the same format as in Fig.~\ref{spectr1}.
Fig.~\ref{spectr2}a is obtained for the times $t_\omega$ and $t_\omega+\Delta t$
as above.
Eq.~(\ref{sptrdyn}) is well satisfied in run~2, $|O_{k}|/Q_{\mu}<0.005$.
Fig.~\ref{spectr2}a shows that the region dominated
by dissipation is wider compared to run~2, owing to the
larger viscosity. The energy containing region
as well as the region where the energy transfer dominates
are shifted to larger scales. 
{ Fig.~\ref{spectr2}a demonstrates
the cumulative behavior of the low-pass filter 
in the $k$ space, Eq.~(\ref{stcum}). 
The pressure-dilatation term is stronger
compared to that in run~1 and reaches the largest value
on large $k$ (small scales). }
The error of the incompressible approach $O_{k}^{(i)}$ 
is larger and is not only connected with
the pressure dilatation; the incompressible
terms (especially the cascade one)
noticeably differ from the compressible ones.

Fig.~\ref{spectr2}b shows 
that over the pressure-dilatation period 
the different components,
$-\partial E_{\mathrm{k}k}/{\partial t}$,
$-S_k$, $ \Psi_k$, and $-D_{k}$ 
have very large temporal variations (large differences
between the minimum and maximum values given by the dotted lines in Fig.~\ref{spectr2}b).  
The averaged pressure dilatation term is weak
and exhibits small negative values over
medium scales, a behavior qualitatively similar to that of
the transfer $-S_k$. This 
indicates that 
the averaged effect of the pressure dilatation 
is a spectral transfer of the kinetic energy
to smaller scales without a net exchange 
between kinetic and internal energies.

The observed spectra in
Fig.~\ref{spec} can be now interpreted using
Figs.~\ref{spectr1} and \ref{spectr2}.
The regions, where the compensated power spectra
$P k^{5/3}$ are about flat, correspond
to regions where the energy transfer rate $S_k$ dominates.
Large scales are dominated by the decay of $E_{\mathrm{k}k}$
and smaller scales are dominated by the dissipation.

\section{K\'arm\'an-Howarth-Monin equation}
\label{cascade}
\subsection{Incompressible HD}

In the incompressible HD 
(see Eq.~(\ref{ivelocity}))
the structure function 
\begin{align} 
\mathcal{S}^{(i)}(\boldsymbol{l})=\langle|\delta\boldsymbol{u}|^{2}\rangle,
\end{align}
(where $\delta\boldsymbol{u}=\boldsymbol{u}(\boldsymbol{x}^\prime)-\boldsymbol{u}(\boldsymbol{x})$,
$\boldsymbol{x}^\prime=\boldsymbol{x}+\boldsymbol{l}$,
and $\langle \bullet \rangle$ denotes spatial averaging) 
describes the kinetic-energy (per mass) spatial scale distribution
and is related to the kinetic-energy power spectrum \cite{fris95}.
For statistically homogeneous decaying turbulence
one can get the following dynamic KHM equation for
the $\mathcal{S}^{(i)}$ \citep{kaho38,moya75}
\begin{equation}
\frac{\partial \mathcal{S}^{(i)}}{\partial t}+
\boldsymbol{\nabla}_{\boldsymbol{l}}\cdot \boldsymbol{\mathcal{Y}}^{(i)}
 =  2 \nu \Delta_{\boldsymbol{l}} \mathcal{S}^{(i)}    - 4 \epsilon,
\label{KHM}
\end{equation}
where 
\begin{align}
\boldsymbol{\mathcal{Y}}^{(i)}(\boldsymbol{l})=\left\langle \delta\boldsymbol{u}|\delta\boldsymbol{u}|^{2}\right\rangle;
\end{align}
{ $\epsilon$ is the incompressible heating rate (see Eq.~\ref{icons}).}
Eq.~(\ref{KHM}) is simply related to its original form that involves 
the cross-correlation $R(\boldsymbol{l})=\left\langle\boldsymbol{u}(\boldsymbol{x}^\prime)
 \cdot\boldsymbol{u}(\boldsymbol{x})\right\rangle$
 \cite{fris95}
\begin{align}
2\frac{\partial R(\boldsymbol{l}) }{\partial t}
-\boldsymbol{\nabla}_{\boldsymbol{l}}\cdot\boldsymbol{\mathcal{Y}}^{(i)} =4\nu\Delta_{\boldsymbol{l}}
R(\boldsymbol{l})
\end{align}
since $S^{(i)}=2\langle |\boldsymbol{u}|^2\rangle -2 R(\boldsymbol{l})$
and $\partial \langle |\boldsymbol{u}|^2\rangle/\partial t=-2\epsilon$.
Eq.~(\ref{KHM}) relates the change of the second order structure function $\mathcal{S}^{(i)}$,
$\partial \mathcal{S}^{(i)}/{\partial t}$, the dissipation rate $\epsilon$, the 
cross-scale transfer/cascade rate
$\mathcal{K}^{(i)}=-\boldsymbol{\nabla}_{\boldsymbol{l}} \cdot \boldsymbol{\mathcal{Y}}^{(i)}/4$,
and the dissipation term $\nu \Delta_{\boldsymbol{l}} \mathcal{S}^{(i)}$
(henceforth we drop the $\boldsymbol{l}$ subscript for $\boldsymbol{\nabla}$ and $\Delta$).
The inertial range can be formally defined as the region where the decay and dissipation terms
are negligible so that
\begin{equation}
 \mathcal{K}^{(i)} = \epsilon.
\label{inertial}
\end{equation}
For isotropic media, in the infinite Reynolds number limit, Eq.~(\ref{inertial}) 
leads to the so called  exact (scaling) laws \citep{kolm41b,fris95}.
Eq.~(\ref{KHM}) is more general and may be directly tested 
in numerical simulations \cite{gotoal02}, since
large Reynolds numbers needed for existence of the inertial range
are computationally challenging
 \cite{ishial09}.

\subsection{Compressible HD}

For the compressible Navier-Stokes equations, Eqs.~(\ref{density},\ref{velocity}),
one possibility to describe the scale-distribution of kinetic energy is the structure function
$\mathcal{S}=\left\langle \delta\boldsymbol{u}\cdot\delta\left(\rho\boldsymbol{u}\right)\right\rangle$
 \cite{gaba11}. For the statistically homogeneous system one gets
\begin{align}
\frac{\partial \mathcal{S}}{\partial t}+\boldsymbol{\nabla}\cdot\boldsymbol{\mathcal{Y}}+\mathcal{R} 
= 4 \varPsi - 4 \mathcal{D} +  C_\Psi-C_D,
\label{KHMc}
\end{align}
where
$\boldsymbol{\mathcal{Y}}=\left\langle \delta\boldsymbol{u}\left[\delta\left(\rho\boldsymbol{u}\right)\cdot\delta\boldsymbol{u}\right]\right\rangle$,
$\mathcal{R}=\left\langle \delta\boldsymbol{u} \cdot \left( \theta^\prime \rho \boldsymbol{u} -\theta \rho^{\prime}\boldsymbol{u}^{\prime}\right)\right\rangle $, 
$\varPsi= \left\langle \delta p\delta\theta\right\rangle /2$, and
$\mathcal{D}= \left\langle \delta\boldsymbol{\tau}:\delta\boldsymbol{\Sigma}\right\rangle/2$.

Here $C_{\varPsi}$ and $C_{\mathcal{D}}$ are correction terms to $\varPsi$ and $\mathcal{D}$ (that 
we choose to represent the pressure dilatation and the dissipation),
respectively,
\begin{align}
C_{\varPsi}= \mathcal{C}_\rho\left[\boldsymbol{u},\boldsymbol{\nabla}p \right], \ \ \ 
C_{\mathcal{D}}=\mathcal{C}_\rho\left[\boldsymbol{u},\boldsymbol{\nabla}\cdot\boldsymbol{\tau}\right], 
\label{correction}
\end{align}
where
\begin{align}
\mathcal{C}_\rho\left[\boldsymbol{a},\boldsymbol{b}\right]
&=\left(\frac{\rho^{\prime}}{\rho}-1\right)\boldsymbol{a}^{\prime}\cdot\boldsymbol{b}+\left(\frac{\rho}{\rho^{\prime}}-1\right)\boldsymbol{a}\cdot\boldsymbol{b}^{\prime}.\nonumber
\end{align}
Note that, the $C_{\varPsi}$ and $C_{\mathcal{D}}$ terms depend explicitly on the level of density fluctuations in the system.

$\mathcal{S}$ and $\boldsymbol{\mathcal{Y}}$ are compressible generalizations 
of $\mathcal{S}^{(i)}$ and $\boldsymbol{\mathcal{Y}}^{(i)}$, respectively.
The $\mathcal{R}$ term presents an additional
compressible energy-transfer channel \cite{gaba11} and likely corresponds
to the compressible part in the spectral transfer, Eq.~(\ref{stsk});
we do not see an obvious way how to turn this term to a divergence form
similar to $\boldsymbol{\nabla}\cdot\boldsymbol{\mathcal{Y}}$. The term
$\varPsi$ is a structure-function formulation of the pressure dilatation effect 
$p \theta$. The viscous term  $\mathcal{D}$  corresponds to a combination of the two
dissipation terms in the incompressible case,
$\epsilon - \nu \Delta \mathcal{S}^{(i)}/2$, in Eq.~(\ref{KHM}).
On large scales, $|\delta \boldsymbol{x}|\rightarrow \infty$, the correlations 
$\left\langle {\boldsymbol{\tau}(\boldsymbol{x}^\prime)}:\boldsymbol{\Sigma} \right\rangle\rightarrow 0$,
and
the viscous term becomes the viscous heating rate $Q_{\mu}$,
\begin{align*}
\mathcal{D} \rightarrow \left\langle \boldsymbol{\tau}:\boldsymbol{\Sigma}\right\rangle = Q_{\mu}.
\end{align*}

The inertial range  may be defined as the interval in the space of separation scales $\boldsymbol{l}$
where
\begin{align}
\mathcal{K}= Q_\mu 
\label{exactKHM}
\end{align}
where $\mathcal{K}=- (\boldsymbol{\nabla}\cdot\boldsymbol{\mathcal{Y}}+\mathcal{R})/4$
is the cascade/energy transfer term.
Eq.~(\ref{exactKHM}) corresponds to the ST relation (Eq.~(\ref{exactST}))
and also neglects the pressure-dilatation effect.
Now we can use Eq.~(\ref{KHMc}) to interpret the simulation results.
We define the departure from zero of this equation as
\begin{align}
\mathcal{O}(l)  &= -\frac{1}{4} \frac{\partial \mathcal{S}}{\partial t} +\mathcal{K} 
 +  \varPsi^\prime  -   \mathcal{D}^\prime .
\label{KHMO} 
\end{align}
where the correction terms were included in $\varPsi^\prime$ and $\mathcal{D}^\prime$, 
$\varPsi^\prime=\varPsi+C_{\varPsi}/2$ and $\mathcal{D}^\prime=\mathcal{D}+C_\mathcal{D}/2$.

\begin{figure}
\centerline{\includegraphics[width=8cm]{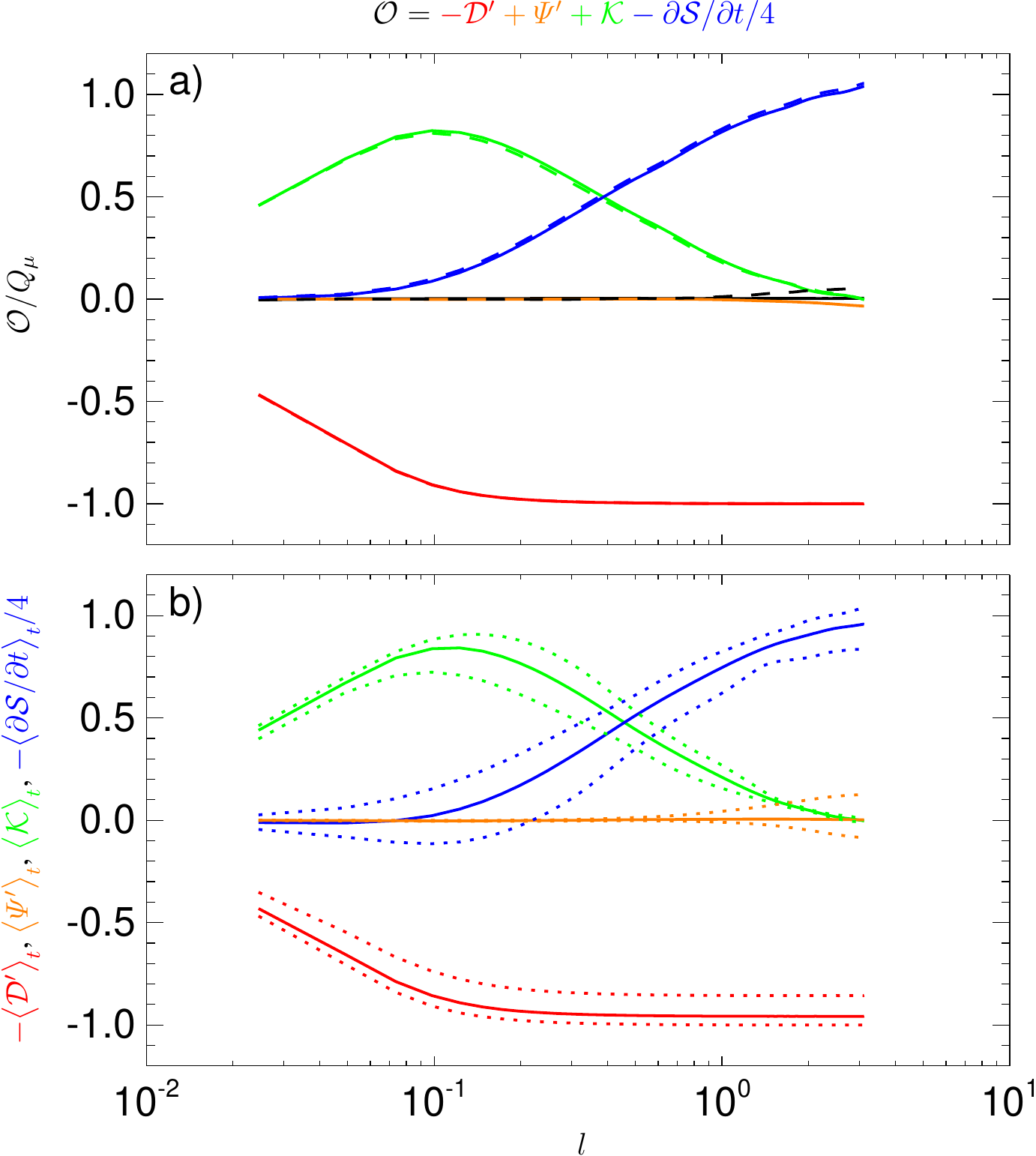}}
\caption{KHM equation for run~1: (a) The validity test
 $\mathcal{O}$ of Eq.~(\ref{KHMO}) (black solid line) as a function of the separation
 scale $l$ along with
the different contributions, (blue) $-{\partial \mathcal{S}}/{\partial t}/4$,
(green) $\mathcal{K}$, (orange) $\Psi^\prime$, and (red) $-D^\prime$.
Dashed lines show the incompressible equivalents.
(b) Time-averaged contributing terms (solid lines)
with their minimum and maximum values (dotted lines) for
 (blue) $-\langle{\partial \mathcal{S}}/{\partial t}\rangle_t/4$,
(green) $\langle\mathcal{K}\rangle_t/4$, (orange) 
$\langle\varPsi^\prime\rangle_t$, and (red)  $-\langle\mathcal{D}^\prime\rangle_t$.
All the quantities are normalized with respect to $Q_{\mu}$.
\label{yag1}
}
\end{figure}

The calculation of structure functions in 3D is computationally
demanding thus the KHM analysis is done on a $256^3$ box
(taking every fourth point in all directions).
The structure functions are calculated over the 3D separation space and
isotropized/averaged over the solid angle. The partial time
derivative is approximated by the finite difference between the two times.
Fig.~\ref{yag1}a shows the validity test
 $\mathcal{O}$ in run~1  as a function of the 
scale $l=|\boldsymbol{l}|$ along with
the different contributions, the decay term $-{\partial \mathcal{S}}/{\partial t}/4$,
the energy transfer/cascade term $\mathcal{K}$,
the pressure dilatation term  $\varPsi^\prime$,
and the dissipation term $ - \mathcal{D}^\prime$.
Eq.~(\ref{KHMc}) is well satisfied in run~1, the departure from validity is small,
$|\mathcal{O}|/Q_\mu<0.005$; this error 
is due to the finite difference
estimation of $\partial \mathcal{S}/\partial t$ (as in
the spectral transfer case).

On large scales, the compressible dissipation term
$\mathcal{D} \rightarrow  Q_\mu $ as expected.
Similarly, $\partial \mathcal{S}/\partial t /4 \rightarrow 
\partial E_\mathrm{k} /\partial t \sim - Q_\mu $.
The pressure-dilatation term is small and appears only on large scales.
The cascade term is important on medium scales but
there is no true inertial range, since both the decay and the dissipation 
are not negligible there.
Run~1 is weakly compressible, the compressible energy-transfer term $\mathcal{R}$
is small ($|\mathcal{R}|/Q_\mu<0.03$). Also the correction terms are negligible
 ($|C_\varPsi|/Q_\mu<0.006$ and $|C_\mathcal{D}|/Q_\mu<0.002$). 

Fig.~\ref{yag1}a displays by dashed lines results of the corresponding
incompressible version of KHM equation, 
the validity test
given by
\begin{equation}
 \mathcal{O}^{(i)}(l)= -\frac{1}{4} \frac{\partial \mathcal{S}^{(i)}}{\partial t}
+ \mathcal{K}^{(i)}
 +  \frac{1}{2} \nu \Delta \mathcal{S}^{(i)}  - \epsilon .
\label{KHMiO}
\end{equation}
The incompressible terms are comparable to their compressible
counterparts, in particular, the dissipation terms are close
to each other since the dissipation is mostly
incompressible (see Fig.~\ref{evolr1}e).
The incompressible error $\mathcal{O}^{(i)}$ appears
on large scales and is related to the missing
pressure-dilatation term, in agreement
with the ST results.

Fig.~\ref{yag1}b displays the results of the KHM equation averaged
over one pressure-dilatation oscillation period.
The colored solid lines show the time averaged quantities,
the decay term $-\partial \mathcal{S}/\partial t/4$,
the energy transfer/cascade term $\mathcal{K}$,
the pressure dilatation term  $\varPsi^\prime$,
and the scale-dependent dissipation term $ - \mathcal{D}^\prime$.
The colored dotted lines shows the corresponding minimum
and maximum values. 
The averaged pressure-dilatation effect is negligible
even though the variation is of the order $0.1 Q_\mu$. 
Similar variations are seen also in other terms.

\begin{figure}
\centerline{\includegraphics[width=8cm]{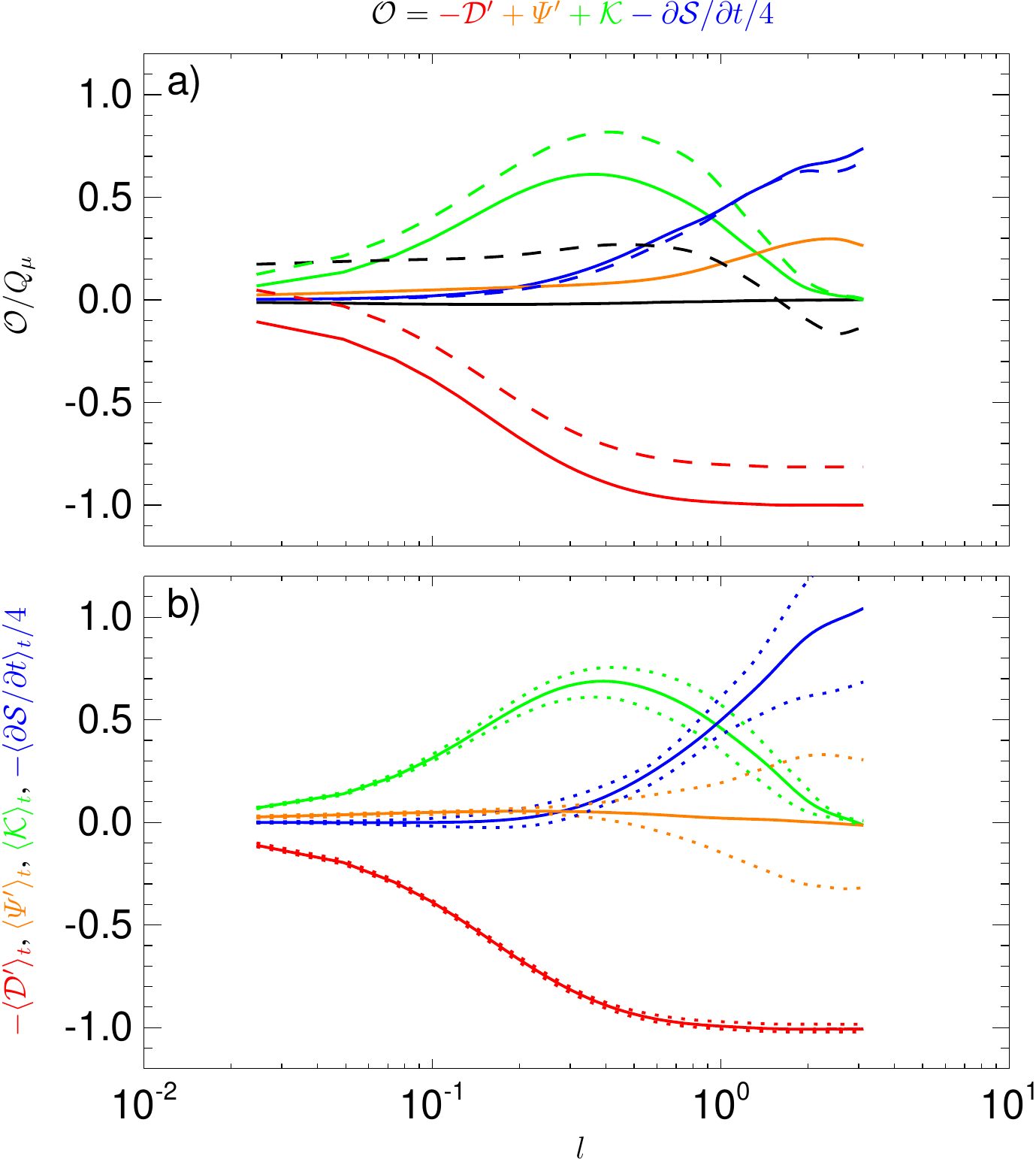}}
\caption{KHM equation for run~2 in the same format 
as in Fig.~\ref{yag1}.
\label{yag2}
}
\end{figure}

The results for the more compressible run~2
are shown in Fig.~\ref{yag2}.
Fig.~\ref{yag2}a displays by solid lines the results for the time $t_\omega$
and $t_\omega + \Delta t$, the error check
 $\mathcal{O}$ as a function of $l$ along with
the different contributions.
Eq.~(\ref{KHMc}) is also well satisfied in run~2, the departure from validity is small,
$|\mathcal{O}|/Q_\mu<0.01$. 
{ The KHM approach exhibits cumulative properties similar
to those of the ST approach (Eq.~(\ref{stcum})).}
The compressible dissipation term 
$\mathcal{D} \rightarrow  Q_\mu $ on large scales as expected,
Similarly, $\partial \mathcal{S}/\partial t /4 \rightarrow
\partial E_\mathrm{k} /\partial t \simeq - 0.7 Q_\mu $
and $\varPsi^\prime  \rightarrow \langle p \theta \rangle
\simeq 0.3 Q_\mu$. 
On large scales we recover the energy conservation
$\partial E_\mathrm{k} /\partial t =
\langle p \theta \rangle -Q_\mu$.

Dashed lines on Fig.~\ref{yag2}a show the incompressible results, 
 $\mathcal{O}^{(i)}$ and its constituents. 
The incompressible KHM equation is not applicable in run~2:
the error $\mathcal{O}^{(i)}$
is substantial and is not simply related to
the pressure dilatation, an important
part of dissipation is compressible (see Fig.~\ref{evolr2}e), and 
the incompressible transfer rate, $\mathcal{K}^{(i)}$,
departs strongly from the compressible
one $\mathcal{K}$. This is partly due to 
the compressible energy-transfer term $\mathcal{R}$
that becomes important ($|\mathcal{R}|/Q_\mu<0.7$).
In run~2, the correction terms are not negligible
 ($|C_\varPsi|/Q_\mu<0.3$ and $|C_\mathcal{D}|/Q_\mu<0.1$).

It is interesting that for run~2 the incompressible approximation
overestimates the energy-transfer rate in
the KHM approach whereas for the ST method
the incompressible equation gives
an energy-transfer rate that is lower
than the compressible one (see Fig.~\ref{spectr2}a).
The incompressible approximation is not generally valid,
however, it may possibly be useful to locate the inertial range.

The different contributing terms 
 of Eq.~(\ref{KHMc}) time-averaged over one pressure-dilatation
oscillation are shown in Fig.~\ref{yag2}b.
All the quantities (except the dissipation one) 
exhibit large variations, mainly on
large scales. The averaged pressure-dilatation term, 
$\langle\varPsi^\prime\rangle_t$, is about zero
on large and small scales and reaches
the maximum $\simeq 0.05 Q_\mu$
at about $l=20$.
The positive value of $\langle\varPsi^\prime\rangle_t$
suggests that the pressure-dilatation effect leads to
a transfer of the kinetic energy from large to smaller scales,
while there is no net energy exchange
between the kinetic and internal energies,
in agreement with the ST results.

Note that the choice  $\delta (\rho \boldsymbol{u})\cdot \delta \boldsymbol{u}$ corresponds
in the ST approach to $\Re \widehat{\rho\boldsymbol{u}}\cdot \widehat{\boldsymbol{u}}^*$ \cite{grahal10}.
 For the ST equation with $|\widehat{\boldsymbol{w}}|^2$ 
 \cite{scgr19,prgi19} (see Eq.~(\ref{sptrdyn}))
 one can obtain an alternative KHM equation taking $\mathcal{S}_w=\left\langle |\delta \boldsymbol{w}|^2\right\rangle$
as
\begin{align}
\frac{\partial \mathcal{S}_w}{\partial t}+\boldsymbol{\nabla}_{\boldsymbol{l}}\cdot\boldsymbol{\mathcal{Y}}_w+\mathcal{R}_w
= 4 \varPsi - 4 \mathcal{D} +  C_{\varPsi w}-C_{\mathcal{D}w},
\label{KHMw}
\end{align}
where 
\begin{align}
\boldsymbol{\mathcal{Y}}_w=\left\langle \delta\boldsymbol{u} |\delta \boldsymbol{w}|^2\right\rangle, \ \ \
\mathcal{R}_w=\left\langle \delta\boldsymbol{w}\cdot\left(\theta^{\prime}\boldsymbol{w}-\theta\boldsymbol{w}^{\prime}\right)\right\rangle
\end{align}
\begin{align}
C_{\varPsi w}= 2\mathcal{C}_{\sqrt{\rho}}\left[\boldsymbol{u},\boldsymbol{\nabla}p \right], 
\ \ \ C_{\mathcal{D}w}=2\mathcal{C}_{\sqrt{\rho}}\left[\boldsymbol{u},\boldsymbol{\nabla}\cdot\boldsymbol{\tau}\right], 
\label{wcorrection}
\end{align}
For the two weakly compressible runs presented here 
these two variants of the KHM relation give almost identical results.

\subsection{Comparison}

For both the runs, the ST and KHM equations give quantitatively analogous results.
This is not surprising, 
$E_{\mathrm{k}k}$ represents a low-pass filtered
spectral distribution of the kinetic energy whereas $\mathcal{S}$ represents
the kinetic energy at the separation scales smaller than $l$ (corresponding to a high-pass filter),
and similar differences apply to the other terms. A remaining question is
the relationship between the wave vector $k$ and the scale separation $l$.
Since the two quantities should be inversely proportional,
we tested different factors $\alpha$ in $k=\alpha/l$. 
For $\alpha=\sqrt{3}$ the ST and KHM results
get close to each other.
\begin{figure}
\centerline{\includegraphics[width=8cm]{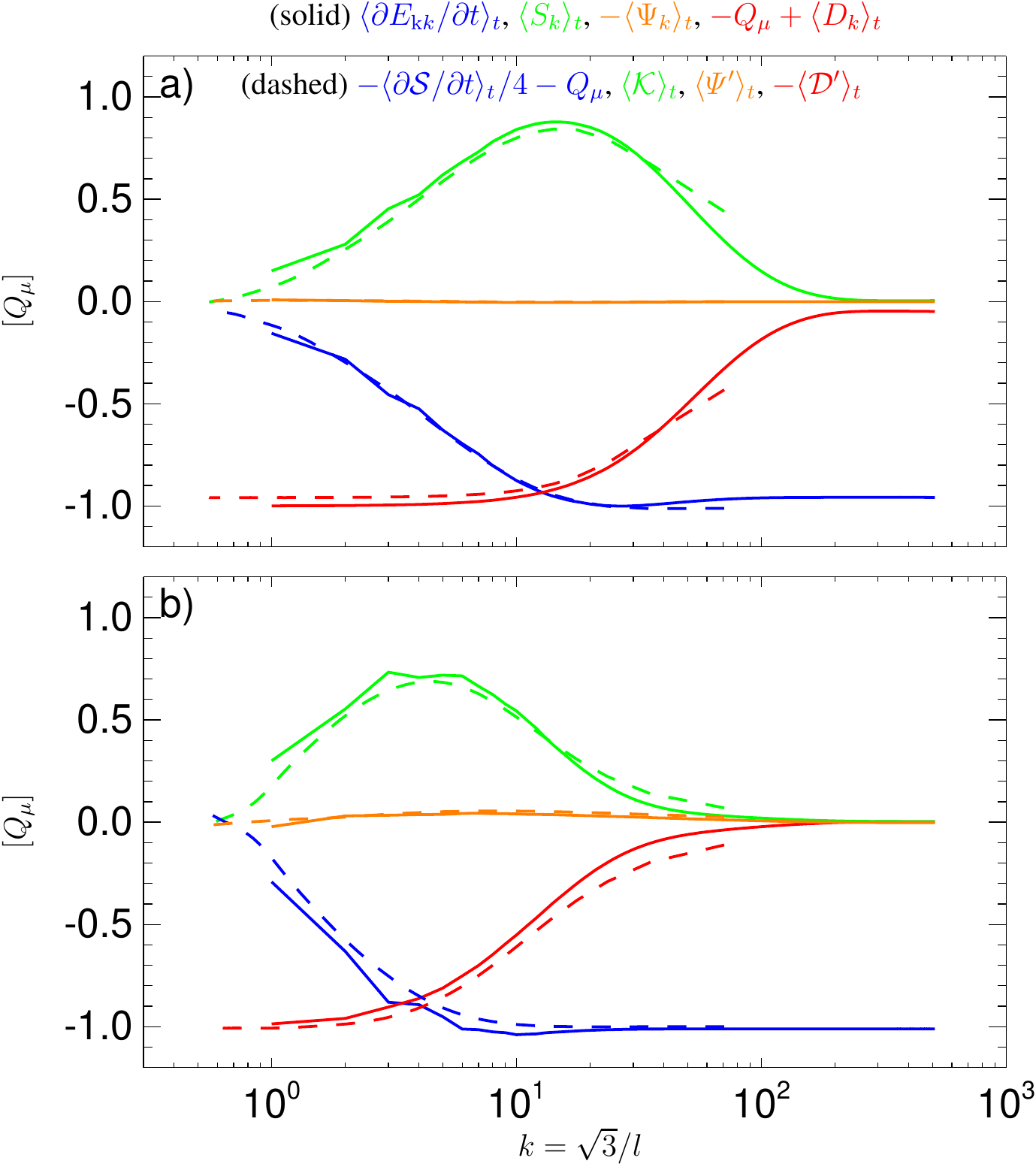}}
\caption{Direct comparison between ST and KHM methods 
for (a) run~1 and (b) run~2:
Solid lines denote the time-averaged ST terms (see Figs.~\ref{spectr1} and~\ref{spectr2}):
(blue) the losses/decay $\langle\partial E_{\mathrm{k}k}/{\partial t}\rangle_t$,
(green) the transfer $\langle S_k\rangle_t$, (orange) the pressure dilatation term 
 $ -\langle\Psi_k\rangle_t$, and (red) 
the dissipation term $ \langle D_{k}\rangle_t -Q_\mu $ (i.e., shifted by $Q_\mu$)
as functions of $k$.
Dashed lines denote the corresponding
time-averaged KHM contributions (see Figs.~\ref{yag1} and~\ref{yag2}):  
(blue) $-\langle{\partial \mathcal{S}}/{\partial t}\rangle_t/4 -Q_\mu$ (i.e., shifted by $Q_\mu$),
(green) $\mathcal{K}$, (orange)
$\langle\Psi^\prime\rangle_t$, and
(red)  $-\langle\mathcal{D}^\prime\rangle_t$ as functions of $l=\sqrt{3}/k$. 
All the quantities are normalized with respect to $Q_{\mu}$.
\label{styag2}
}
\end{figure}
Fig.~\ref{styag2} shows that in both runs,
for $k=\sqrt{3}/l$, the time-averaged cascade rates
obtained from the ST and KHM relations are comparable
$\langle S_k\rangle_t \simeq \langle\mathcal{K}\rangle_t$;
the same is true for the pressure-dilatation induced cross-scale transfers
$\langle\Psi_k\rangle_t \simeq -\langle\varPsi\rangle_t$.

The decay and dissipation terms have comparable behaviors
when shifted by the dissipation rate $Q_\mu$. This
may be expressed as (here we leave out the time averages) 
\begin{align}
 \frac{\partial E_{\mathrm{k}k}}{\partial t}
+ \frac{1}{4} \frac{\partial \mathcal{S}}{\partial t} & \simeq  -Q_\mu, \ \ \
 D_{k} + \mathcal{D}^\prime   \simeq  Q_\mu,
\end{align}
the ST and KHM quantities are complementary as expected.
As $\partial E_{\mathrm{k}k}/{\partial t}$ represents
the rate of change of the kinetic energy on scales
with wave-vector magnitudes smaller or equal to $k$, 
${\partial \mathcal{S}}/{\partial t}/4$ gives
approximatively the remaining decay rate 
(for wave-vector magnitudes larger than $k$).
Similarly, $D_{k}$ is the dissipation rate
on the scales $\le k$ whereas $\mathcal{D}^\prime$
represents about the complementary dissipation rate
(on the scales $> k$).

\section{Internal Energy}

\label{inten}

In the previous section, we showed
that the exchanges
between the kinetic and internal energies lead to
a transfer of kinetic energy from large to
small scales. It is, therefore, interesting to
look at the scale dependence of the internal
energy and its cross-scale transfer.  

\subsection{Spectral transfer}

One possible description of the spectral
scale decomposition and cross-scale transfer of the
internal energy could be done through 
the variable $q=(\gamma \rho T)^{1/2}$ 
\cite{scgr19}. Its evolution, following from Eq.~(\ref{temperature}),
is given by
\begin{align}
\frac{\partial q}{\partial t}+\left(\boldsymbol{u}\cdot\boldsymbol{\nabla}\right)q+\frac{1}{2}q\theta=&\frac{\alpha\gamma}{2}\frac{1}{q}\Delta T-\frac{1}{2}\gamma\left(\gamma-1\right)\frac{1}{q}p\theta \\
& +\frac{1}{2}\gamma\left(\gamma-1\right)\frac{1}{q}\boldsymbol{\Sigma}:\boldsymbol{\tau}. \nonumber
\end{align}
We set the spectral decomposition of the internal energy,
analogously to the case of the kinetic one, as
low-pass filtered quantity
\begin{align}
E_{\mathrm{i}k}=
\frac{1}{\gamma\left(\gamma-1\right)}
 \sum_{|\boldsymbol{k}^\prime|\le k}
|\widehat{q}|^{2}.
\end{align}
For $E_{\mathrm{i}k}$, one gets the following dynamic equation
\begin{align}
\frac{\partial E_{\mathrm{i}k}}{\partial t} +  S_{\mathrm{i}k} = \Phi_{\mathrm{i}k} 
-\Psi_{\mathrm{i}k} + D_{\mathrm{i}k}
\label{esptrdyn}
\end{align}
where 
\begin{align}
S_{\mathrm{i}k}&=\frac{1}{\gamma\left(\gamma-1\right)} \Re \sum_{|\boldsymbol{k}^\prime|\le k}
\left[
2\widehat{q}^{*}\widehat{\left(\boldsymbol{u}\cdot\boldsymbol{\nabla}\right)q}+\widehat{q}^{*}\widehat{q\theta}
\right], \nonumber \\
\Psi_{\mathrm{i}k}&=
\frac{1}{\gamma} \Re \sum_{|\boldsymbol{k}^\prime|\le k}  \widehat{q}^{*}\widehat{q\theta}, \ \ \
D_{\mathrm{i}k} = \Re \sum_{|\boldsymbol{k}^\prime|\le k}
\widehat{q}^{*}\widehat{q^{-1}\boldsymbol{\Sigma}:\boldsymbol{\tau}}, \nonumber \\
\Phi_{\mathrm{i}k}&=
\frac{\alpha}{\left(\gamma-1\right)}
\Re \sum_{|\boldsymbol{k}^\prime|\le k}
\widehat{q}^{*}\widehat{q^{-1}\Delta T}.
\end{align}
Here $S_{\mathrm{i}k}$ describes the cross-scale energy transfer,
$\Phi_{\mathrm{i}k}$ results from the thermal diffusion,
$\Psi_{\mathrm{i}k}$ is a term representing the
pressure-dilatation effect, and $D_{\mathrm{i}k}$ 
comes from the viscous heating.

\begin{figure}
\centerline{\includegraphics[width=8cm]{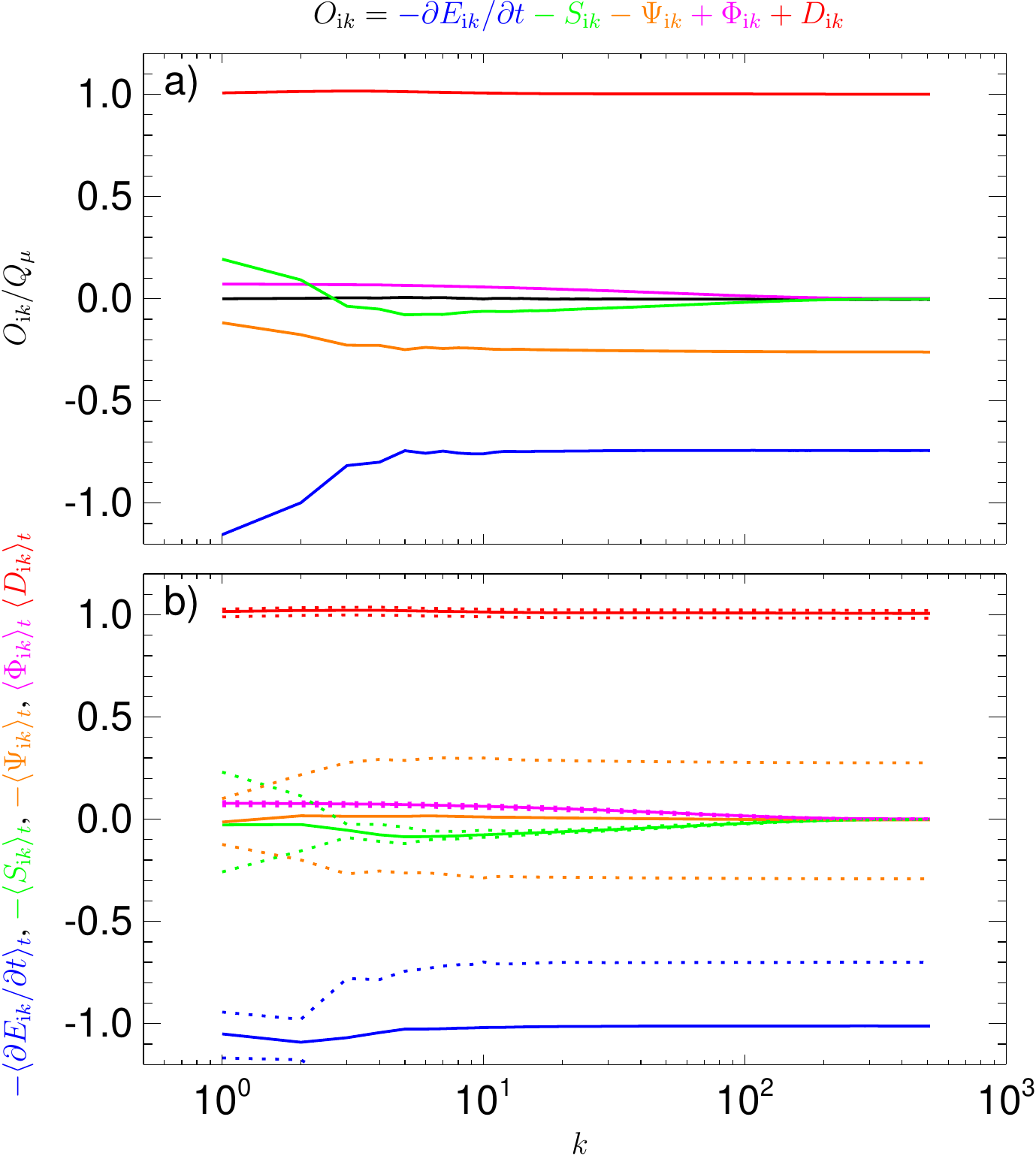}}
\caption{Spectral transfer of the internal energy in run~2: (a) 
 The validity test  $O_{\mathrm{i}k}$ of Eq.~(\ref{oesptrdyn}) (black line)
 as a function of $k$ along with
the different contributions, (blue) the time variation $-\partial E_{\mathrm{i}k}/{\partial t}$,
(green) the energy transfer term $-S_{\mathrm{i}k}$, (orange) the pressure dilatation term
 $ \Psi_{\mathrm{i}k}$,
(magenta) the diffusion $\Phi_{\mathrm{i}k}$,
 and (red) the dissipation term $ D_{\mathrm{i}k} $.
(b) Time-averaged contributing terms (solid lines)
with their minimum and maximum values (dotted lines) for
(blue) the decay $-\langle\partial E_{\mathrm{i}k}/{\partial t}\rangle_t$,
(green) the transfer $-\langle S_{\mathrm{i}k}\rangle_t$, (orange) the pressure dilatation term
 $ \langle\Psi_{\mathrm{i}k}\rangle_t$,
(magenta) the diffusion $\langle\Phi_{\mathrm{i}k}\rangle_t$,
 and (red) the dissipation term $ \langle D_{\mathrm{i}k}\rangle_t $.
All the quantities are normalized with respect to $Q_{\mu}$.
\label{spectrie}
}
\end{figure}

As there is no clear pressure-dilation induced
cross-scale transfer in run~1, we look 
only at run~2.
We define the validity test of Eq.~(\ref{esptrdyn})
as before by
\begin{align}
O_{\mathrm{i}k}= -\frac{\partial E_{\mathrm{i}k}}{\partial t} -  S_{\mathrm{i}k} + \Phi_{\mathrm{i}k}
-\Psi_{\mathrm{i}k} + D_{\mathrm{i}k}.
\label{oesptrdyn}
\end{align}
Fig.~\ref{spectrie}a 
displays $O_{\mathrm{i}k}$, and its constituents, obtained 
at $t_\omega$ and $t_\omega+\Delta t$.
Eq.~(\ref{esptrdyn}) is well satisfied, $|O_{\mathrm{i}k}|/Q_\mu < 0.007$.
$\partial E_{\mathrm{i}k}/{\partial t}$ is positive as the internal energy increases
and varies mostly on large scales. 
$D_{\mathrm{i}k}$ is about constant $\sim Q_\mu$.  This
is due the fact that the nonlinear
term $\boldsymbol{\tau}:\boldsymbol{\Sigma}$ heats everywhere
in the simulation box and importantly contributes to
the $k=0$ term.
The pressure-dilatation term varies on large scales, 
the diffusion and the transfer term lead to weak
scale redistribution of the internal energy.

Fig.~\ref{spectrie}b shows the spectral transfer 
results averaged over one pressure-dilatation 
oscillation period, the mean values of the different terms  
and their minimum and maximum values. 
The dissipation, $\langle D_{\mathrm{i}k}\rangle_t\simeq Q_\mu$
with weak variations, 
$\langle\partial E_{\mathrm{i}k}/{\partial t}\rangle_t \simeq Q_\mu$ 
with large temporal variations. 
The pressure dilatation $\langle \Psi_{\mathrm{i}k} \rangle_t$
is small and negative (with large fluctuations).
The diffusion $\langle \Psi_{\mathrm{i}k} \rangle_t$ 
is weak with positive values and 
the cross-scale transfer $\langle S_{\mathrm{i}k} \rangle_t$
is small with large fluctuations of large scales.  
In analogy with the spectral analysis for the kinetic
energy, the nonlinear term
 $S_{\mathrm{i}k}$ leads to transfer of the internal
energy from large to small scales whereas the diffusion and the pressure dilatation 
lead to transfer of the internal energy in the opposite direction. These
processes roughly compensate each other and the dominant
energy channel is the viscous heating
$\langle\partial E_{\mathrm{i}k}/{\partial t}\rangle_t \simeq \langle D_{\mathrm{i}k}\rangle_t$.
It is also clear that the dynamic spectral description
 of the internal energy,
Eq.~(\ref{esptrdyn}),
is hardly comparable to that for the kinetic energy,
Eq.~(\ref{sptrdyn}),
especially concerning the viscous dissipation,
compare Figs.~\ref{spectr2} and~\ref{spectrie}. On the other hand,
the pressure-dilatation terms for the kinetic (Eq.~(\ref{sptrdyn})) and internal (Eq.~(\ref{esptrdyn})) energies
are comparable, $\Psi_{\mathrm{i}k} \simeq \Psi_{k}$.
For the combined quantity $E_{\mathrm{k}k}+E_{\mathrm{i}k}$ the pressure-dilatation terms
cancel each other as one may expect. On the other hand, the dissipation terms have very different
scale representations, so that $E_{\mathrm{k}k}+E_{\mathrm{i}k}$ clearly does not
represent the total energy, the kinetic and internal energies ought to be treated separately.

\subsection{KHM equation}
\label{apIE}

One way to represent the internal energy in
the KHM approach is the structure function \cite{gaba11}
\begin{align}
\mathcal{S}_{\mathrm{i}}=\left\langle \delta\rho\delta e\right\rangle .
\end{align}
From Eq.~(\ref{temperature}), it follows for $e$
\begin{align}
\rho\frac{\partial e}{\partial t}+\rho (\boldsymbol{u}\cdot\boldsymbol{\nabla})e=
\alpha\Delta e-p\theta
+ \boldsymbol{\tau}: \boldsymbol{\Sigma}
\end{align}
and for $\mathcal{S}_{\mathrm{i}}$ one gets
the dynamic KHM-like equation 
\begin{align}
\frac{\partial \mathcal{S}_{\mathrm{i}}}{\partial t}+\boldsymbol{\nabla}
\cdot\boldsymbol{\mathcal{Y}}_{\mathrm{i}}+\mathcal{R}_{\mathrm{i}}&=
 2\varPhi_{\mathrm{i}} - 2\varPsi_{\mathrm{i}}  + 2\mathcal{D}_{\mathrm{i}} 
\label{eKHM}
\end{align}
where
\begin{align}
\boldsymbol{\mathcal{Y}}_{\mathrm{i}}&=\left\langle \delta\boldsymbol{u}\delta\rho\delta e\right\rangle,
 \ \ \
\mathcal{R}_{\mathrm{i}}=\left\langle (\rho \theta^{\prime} -\rho^{\prime}\theta)\delta e \right\rangle,
\nonumber \\ 
\varPsi_{\mathrm{i}} &= \mathcal{V}_\rho\left(p\theta\right)/2, \ \ \
\mathcal{D}_{\mathrm{i}} = \mathcal{V}_\rho\left(\boldsymbol{\tau}:\boldsymbol{\Sigma}\right)/2\nonumber \\
\varPhi_{\mathrm{i}} & = \alpha\left\langle \delta\rho\delta\left(
\rho^{-1} \Delta e\right)\right\rangle/2,
\end{align}
and 
\begin{align*}
\mathcal{V}_\rho\left(a\right)&=\left\langle \left(1-\frac{\rho}{\rho^{\prime}}\right)a^{\prime}+\left(1-\frac{\rho^{\prime}}{\rho}\right)a\right\rangle 
\end{align*}
In Eq.~(\ref{eKHM}) $\mathcal{K}_\mathrm{i}=-(\boldsymbol{\nabla}\cdot\boldsymbol{\mathcal{Y}}_{\mathrm{i}}+\mathcal{R}_{\mathrm{i}})/2$
represents the cross-scale transfer connected with $\mathcal{S}_{\mathrm{i}}$.

The pressure-dilatation  $\varPsi_{\mathrm{i}}$ and the
dissipation $\mathcal{D}_{\mathrm{i}}$ terms 
depend on the density variation; for a constant $\rho$ these
terms disappear. This is the first indication that
$\mathcal{S}_{\mathrm{i}}$ does not represent the internal energy
in a way comparable to the kinetic energy structure function $\mathcal{S}$.

To test Eq.~(\ref{eKHM}) on the simulation results of run~2
we define the departure as
\begin{align}
\mathcal{O}_{\mathrm{i}}(l)&= -\frac{1}{2}\frac{\partial \mathcal{S}_{\mathrm{i}}}{\partial t}+
\mathcal{K}_\mathrm{i}
+\Phi_{\mathrm{i}} - \Psi_{\mathrm{i}}  + D_{\mathrm{i}} .
\label{oe}
\end{align}
\begin{figure}
\centerline{\includegraphics[width=8cm]{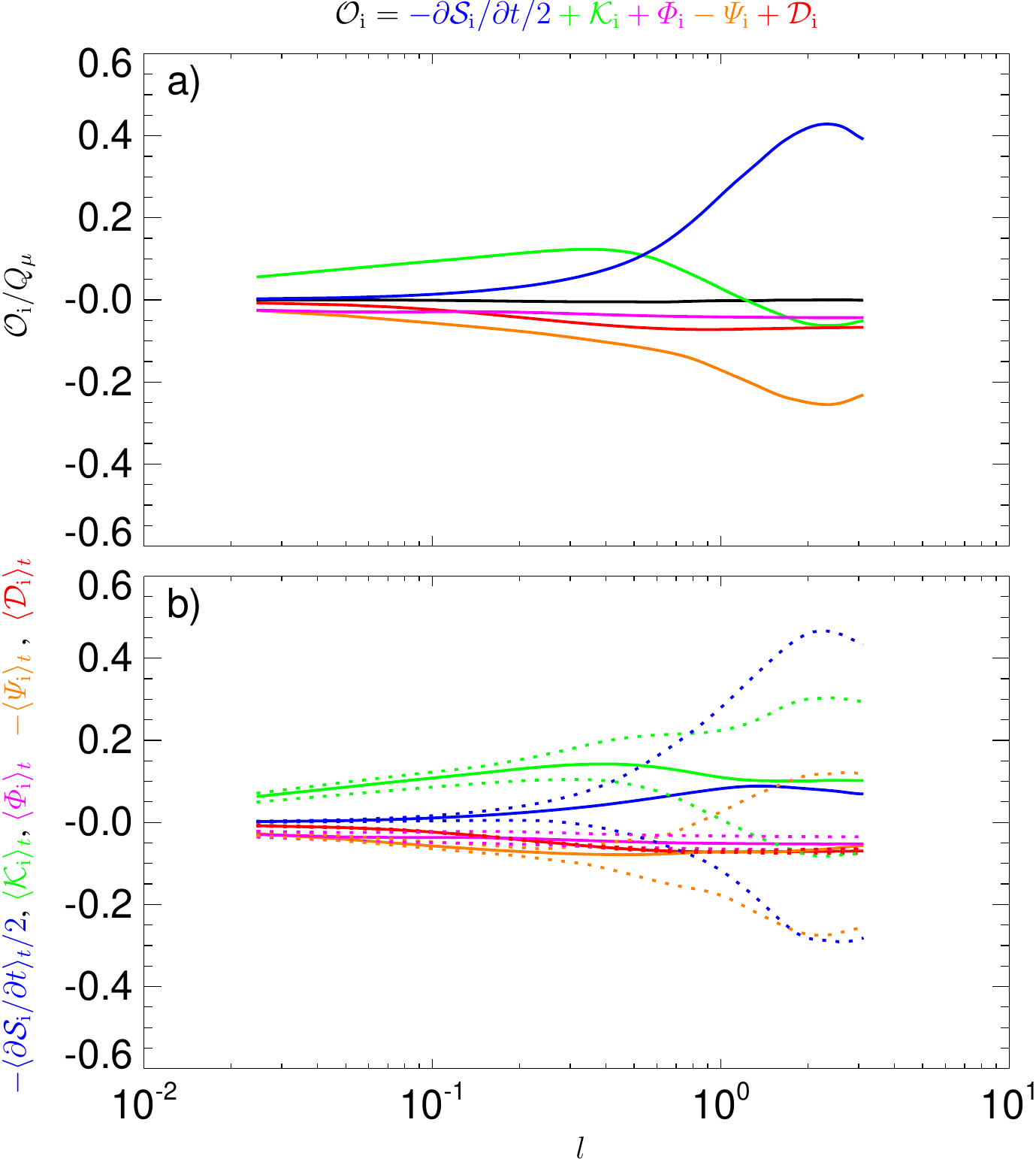}}
\caption{KHM equation for the internal energy
in run~2: (a) The validity test 
 $\mathcal{O}_{\mathrm{i}}$ of Eq.~(\ref{oe}) (black line) 
as a function of the separation scale $l$ along with
the different contributions, (blue) $-{\partial \mathcal{S}_{\mathrm{i}}}/{\partial t}/2$,
(green) $\mathcal{K}_{\mathrm{i}}$, 
(orange) $-\varPsi_\mathrm{i}$, 
(red) $\mathcal{D}_\mathrm{i}$,
and (magenta) $\varPhi_\mathrm{i}$.
(b) Time-averaged contributing terms (solid lines) 
with their minimum and maximum values (dotted lines) for
(blue) $-\langle{\partial \mathcal{S}_{\mathrm{i}}}/{\partial t}\rangle_t/2$,
(green) $\langle \mathcal{K}_{\mathrm{i}}\rangle_t$,
(orange) $-\langle\varPsi_\mathrm{i}\rangle_t$,
(red) $\langle\mathcal{D}_\mathrm{i}\rangle_t$,
and (magenta) $\langle\varPhi_\mathrm{i}\rangle_t$.
All the quantities are normalized with respect to to $Q_{\mu}$.
\label{eyag}
}
\end{figure}
Fig.~\ref{eyag}a shows the departure (black) 
$\mathcal{O}_{\mathrm{i}}$ as a function of the scale $l$ along with
the different contributions, the decay term (blue) $-{\partial \mathcal{S}_{\mathrm{i}}}/{\partial t}/2$,
the energy transfer term (green) $\mathcal{K}_\mathrm{i}$,
the pressure dilatation term (orange) $-\varPsi_\mathrm{i}$,
(red) the dissipation term $\mathcal{D}_\mathrm{i}$,
and the diffusion term (magenta) $\varPhi_\mathrm{i}$.
The calculation is done on a sub-grid of $256^3$.
Eq.~(\ref{eKHM}) is well satisfied, 
$|\mathcal{S}_{\mathrm{i}}|/Q_\mu < 0.005$.

The pressure dilatation structure function terms for the kinetic (Eq.~(\ref{KHMc}))
and internal (Eq.~(\ref{eKHM})) energies are similar, $\varPhi_\mathrm{i} \simeq \varPhi$.
The diffusion term is small and (except for the sign) corresponds 
the diffusion term in the ST approach (Eq.~(\ref{esptrdyn})).
The dissipation terms is small with respect to the
dissipation rate $Q_\mu$, indicating that the viscous heating is 
not well represented in Eq.~(\ref{eKHM}). Consequently,
the structure function $\mathcal{S}_{\mathrm{i}}=\left\langle \delta\rho\delta e\right\rangle$ decreases
with time in contrast with the internal energy $E_\mathrm{i}=\langle \rho e \rangle$
that increases (see Fig.~\ref{evolr2}).
These properties remain unchanged even after averaging over one pressure-dilatation 
oscillation period as displayed in 
Fig.~\ref{eyag}b. 
All the terms (except the dissipation and diffusion ones) exhibit large
variations, dominantly on the large scales. 
The averaged pressure-dilatation term is small and 
corresponds to that of the kinetic energy,
$\langle\varPhi_\mathrm{i}\rangle_t \simeq \langle\varPhi\rangle_t$.

Combining Eq.~(\ref{KHMc})
with Eq.~(\ref{eKHM}) as  $\partial (\mathcal{S}/2+\mathcal{S}_{\mathrm{i}})/\partial t $
one recovers to a large extent the results of Ref.~\onlinecite{gaba11}
(note, however, that the pressure-dilatation effects are in Ref.~\onlinecite{gaba11} transformed to
a contribution to the cascade term using the isothermal closure).
For the combined quantity $\mathcal{S}/2+\mathcal{S}_{\mathrm{i}}$
the pressure-dilation terms cancel each other, similar to the ST case.
However, the scale-dependence of the viscous dissipation/heating is
significantly different in the two approaches, so that it is hard to
interpret $\mathcal{S}/2+\mathcal{S}_{\mathrm{i}}$ as a representative
of the total energy. The kinetic and internal energies are better
to be investigated separately by Eq.~(\ref{KHMc}) and Eq.~(\ref{eKHM}).

The choice $\delta \rho \delta e$ does not correspond to the 
ST equation in the previous section based on $|\widehat{q}|^2$
\cite{scgr19}. In order to get an alternative
version of the internal energy KHM equation corresponding to Eq.~(\ref{esptrdyn}),
one can investigate $|\delta q|^{2}$. The resulting equation 
can be expressed in a form similar to
Eq.~(\ref{eKHM}) as
\begin{align}
\frac{\partial \mathcal{S}_{\mathrm{i}q}}{\partial t}+\boldsymbol{\nabla}
\cdot\boldsymbol{\mathcal{Y}}_{\mathrm{i}q}+\mathcal{R}_{\mathrm{i}q}&=
 \varPhi_{\mathrm{i}q} - \varPsi_{\mathrm{i}q}  + \mathcal{D}_{\mathrm{i}q}
\label{qKHM}
\end{align}
where
\begin{align}
\mathcal{S}_{\mathrm{i}q}&=\frac{1}{\gamma\left(\gamma-1\right)}\left\langle |\delta q|^{2}\right\rangle,
\ \ \
\boldsymbol{\mathcal{Y}}_{\mathrm{i}q}=\frac{1}{\gamma\left(\gamma-1\right)}	\left\langle \delta\boldsymbol{u}|\delta q|^{2}\right\rangle, \nonumber \\
\mathcal{R}_{\mathrm{i}q}	&=\frac{1}{\gamma\left(\gamma-1\right)}\left\langle \delta q (q\theta^{\prime}- q^{\prime}\theta)\right\rangle, \nonumber \\
\varPsi_{\mathrm{i}q}&=\left\langle \delta q\delta\left(q^{-1} p\theta\right)\right\rangle, \ \ \ 
\mathcal{D}_{\mathrm{i}}	=\left\langle \delta q\delta\left(q^{-1}\boldsymbol{\Sigma}:\boldsymbol{\tau}\right)\right\rangle, \nonumber \\
\varPhi_{\mathrm{i}q}	&=\frac{\alpha}{\left(\gamma-1\right)}\left\langle \delta q\delta\left(q^{-1}\Delta T\right)\right\rangle. 
\end{align}
Analysing run~2 using this form of the internal energy KHM equation
we obtain results similar to those in
Fig.~\ref{eyag}. Therefore,
also Eq.~(\ref{qKHM}) is to be investigated separately from Eq.~(\ref{eKHM}).

\section{Discussion}
\label{discussion}
In this paper we investigated the properties
of the spectral/spatial-scale distribution and the cross-scale transfer
of the kinetic energy in compressible hydrodynamic turbulence. 
We used the dynamic spectral transfer (ST) 
 K\'arm\'an-Howarth-Monin (KHM) 
equations, in compressible and incompressible forms,
to analyze results of two  3D direct numerical simulations of 
decaying compressible turbulence simulation with moderate Reynolds numbers and
the initial Mach numbers
  $M=1/3$ and $M=1$.
The simulations are initiated with large-scale solenoidal velocity fluctuations.
The nonlinear coupling leads to a flux of the kinetic energy to small scales where
it is dissipated; at the same time,
the reversible pressure-dilatation mechanism causes oscillatory
exchanges between the kinetic and internal
energies with an average zero net energy transfer.
While the simulations do not exhibit a clear inertial range,
owing largely to moderate Reynolds numbers, 
the dynamic compressible KHM and ST equations are
well satisfied in the simulations. These approaches describe,
in a quantitatively similar way for both the methods, the decay
of the kinetic energy on large scales,
the energy transfer/cascade, the pressure dilatation, and the dissipation process.
The incompressible versions are not valid, especially in run~2 (starting with $M=1$).

{
The ST approach that uses a low-pass filter in the $k$ space is by construction
cumulative; in particular, the dissipation ST term reaches its (absolute)
maximum values (given by the dissipation rate) at large $k$ (small
scales). The KHM approach is complementary and has similar cumulative
properties but in the opposite direction: the dissipation KHM term reach its
(absolute) maximum values at large scales (given as well by the
dissipation rate). The comparison between the two
approaches demonstrates that the range of scales where
the dissipation is important is determined
by the variations/gradients of the ST and KHM dissipation
terms rather than their values.
The same applies to the pressure dilatation:
the pressure-dilatation terms in the ST and KHM
exhibit opposite cumulative properties, they
approaches reach the average pressure dilatation at
small and large scales, respectively.
These results indicate that analyses based on the values of
the cumulative pressure-dilatation terms are not very relevant;
it is the variation over scales that counts.
The pressure-dilatation energy exchange between the kinetic and internal energy
 gets negligible when averaged over a period of pressure-dilatation
oscillations. The time-averaged pressure dilatation may lead to a
transfer of the kinetic energy from large to small scales
(in agreement with Ref.~\onlinecite{wangal18}).
For much larger systems we expect that 
the pressure-dilatation energy exchange becomes negligible for any given time.
This may explain the apparent discrepancy between 
the results of Refs.~\onlinecite{aluial12,prgi19} and Ref.~\onlinecite{wangal18}.
}

The results of both the simulations indicate a simple 
relationship between the KHM and ST results
through the inverse proportionality  between the wave vector $k$
and the spatial separation length $l$ as $k l \simeq \sqrt{3}$
and suggest a complementary scale-distribution meaning of
the ST and KHM quantities.
Interestingly, preliminary results of a similar comparison in two-dimensional
Hall MHD simulations suggest a similar dependence $k l \simeq \sqrt{2}$ indicating
that the relationship between the two scales depends on the space dimension.
{ The simple relationship is useful to interpret the KMH
results in the context of spectral analyses.
}

The ST approach is straightforward, requires less computational resources,
and is directly linked to the spectral properties of velocity
fluctuations. The KHM is more computationally demanding
but leads to the so called exact scaling laws, and can be
directly used to analyze anisotropic turbulence \cite{cambal13,verdal15}.
We obtained similar results from the coarse-graining approach
\cite{hellal20f}. The coarse-graining approach presents 
semi-quantitatively similar results concerning the energy-transfer/cascade, 
decay, dissipation, and the pressure dilatation processes; the localization of
these different processes is, however, somewhat different when expressed 
in space-filtering scales with respect to the spatial separation scale.
{
The cumulative features of the coarse-graining approach is similar
to that of the KHM equation by construction (spatial low-pass filter) but
more detailed comparison between the
coarse-graining method and the ST and KHM ones is beyond the
scope of this paper. }

We also investigated the properties of the internal energy
using dynamic ST and KHM equations. These equations
are well satisfied in both the simulations and
the descriptions of the pressure-dilation effect are
compatible with their counterpart for the kinetic energy.
The ST and KHM equations for the kinetic and internal energies behave,
however,  very differently with respect to the viscous dissipation. 
Consequently, the ST and KHM (and likely also coarse-graining) approaches should better be used 
for the kinetic and internal energies separately. Moreover,
the pressure-dilatation reversible coupling does not appear
to lead to a net energy transfer between the kinetic
and internal energies, at least in weakly compressible
systems. It is, therefore, not necessary to investigate
the two energies combined.
The usage of combined quantities \cite{gaba11,baga14} may lead to questionable
results. For instance, in order to determine 
heating rates of the turbulent cascade it is necessary
to look at the behavior of the kinetic energy 
(plus the magnetic energy in the magnetohydrodynamic case);
the cascade/cross-scale transfer of the internal
energy just leads to its redistribution.

{
Ref.~\onlinecite{aluial12} analyzed the pressure-dilatation 
effect and showed it decreases
rapidly from large to small scales so that for a large enough system
there are scales where the pressure dilatation becomes negligible and where
 the kinetic energy cascades in a conservative manner owing to the nonlinear advection term.
Our results further suggest that on larger scales 
the kinetic energy is also conservatively transferred
from large to small scales, partly owing to the standard nonlinear-advection cascade 
and partly to the pressure-dilatation-induced energy transfer
(the locality of the latter process is unclear).
Our simulation results are limited by moderate Reynolds numbers
 and weak compressibilities, so they need to be extended to larger
Reynolds number and higher Mach numbers
\cite{ishial09,eydr18,drey18,krital13,ferral20}.
}

\acknowledgments
PH acknowledges grant 18-08861S of the Czech Science Foundation.

\end{document}